%% file: main.tex
\documentclass[11pt,a4paper]{article}

\pdfoutput=1
\usepackage{amssymb}
\usepackage{amsmath}
\usepackage{graphicx}
\usepackage{booktabs}
\usepackage{color}
\usepackage{textcomp}
\usepackage{multirow}
\usepackage{bm}
\usepackage{caption}
\usepackage{subcaption}
\usepackage{tikz}
\usepackage{pgf}
\usepackage{pgfkeys}
\usepackage{braket}
\usepackage{longtable}
\usepackage{colortbl}
\usepackage{rotating}
\usepackage{jheppub}   
\usepackage{soul}
\usepackage[utf8]{inputenc}

\usepackage{flafter}

\definecolor{jlab_red}{RGB}{192,39,45}
\definecolor{jlab_orange}{RGB}{249,102,0}
\definecolor{jlab_blue}{RGB}{47,122,121}
\definecolor{jlab_green}{RGB}{65,125,10}

\newcommand{\estate}[1]{\ensuremath{\mathfrak{#1}}}

\newcommand{\Ndf}{\ensuremath{N_\mathrm{dof}}}

\newcommand{\Jpsi}{J/\psi}
\newcommand{\pprime}{\vec{p}^{\, \prime}}

 \hypersetup{%
 pdftitle = {Radiative Transitions in Charmonium from Lattice QCD},
 pdfauthor = {Hadron Spectrum Collaboration},
 colorlinks = {true},
}{}

\begin{document}

\input{title}

\maketitle

\input{introduction}

\input{transitions}

\input{two_points}

\input{three_points}

\input{calc_details}

\clearpage

\input{misc_results}
\clearpage
\input{etace_etace}
\input{etace_prime_etace_prime}
\input{chice_chice}
\clearpage
\input{psice_etace}

\clearpage
\input{chice_psice}

\clearpage
\input{psice_psice}
\clearpage

\input{conclusion}

\bigskip

\begin{acknowledgments}
	We thank our colleagues within the Hadron Spectrum Collaboration (\url{www.hadspec.org}) and in particular Cian O'Hara who carried out some earlier preliminary analyses~\cite{OHara:2017eoq,OHara:2019oeb}.
	CET and JD acknowledge support from the U.K.\ Science and Technology Facilities Council (STFC) [grant number ST/T000694/1].
	SMR acknowledges support from Science Foundation Ireland [grant number 21/FFPP/10186], the Irish Research Council [grant number GOIPG/2016/1245] and
	the European Community Horizon 2020 -- research and innovation programme, contract STRONG-2020 grant number 824093.

	This work used the Seagull cluster maintained by the Trinity Centre for High Performance Computing. This work also used the DiRAC Data Analytic system at the University of Cambridge, operated by the University of Cambridge High Performance Computing Service on behalf of the STFC DiRAC HPC Facility (www.dirac.ac.uk). This equipment was funded by a BIS National E-infrastructure capital grant [ST/K001590/1], STFC capital grants [ST/H008861/1 and ST/H00887X/1], and a STFC DiRAC Operations grant [ST/K00333X/1]. DiRAC is part of the National E-Infrastructure.
	\texttt{iMinuit} \cite{iminuit} was used for the minimization of the $\chi^2$ function.
	Some software codes used in this project were developed with support from the U.S.\ Department of Energy, Office of Science, Office of Advanced Scientific Computing Research and Office of Nuclear Physics, Scientific Discovery through Advanced Computing (SciDAC) program; also acknowledged is support from the Exascale Computing Project (17-SC-20-SC), a collaborative effort of the U.S.\ Department of Energy Office of Science and the National Nuclear Security Administration.

	The software codes {\tt Chroma}~\cite{Edwards:2004sx} and {\tt QUDA}~\cite{Clark:2009wm,Babich:2010mu} were used to compute the propagators required for this project on clusters at Jefferson Laboratory under the USQCD Initiative and the LQCD ARRA project.
	Gauge configurations were generated using an award of computer time provided by the U.S.\ Department of Energy INCITE program and supported in part under an ALCC award, and resources at: the Oak Ridge Leadership Computing Facility, which is a DOE Office of Science User Facility supported under Contract DE-AC05-00OR22725; the National Energy Research Scientific Computing Center (NERSC), a U.S.\ Department of Energy Office of Science User Facility located at Lawrence Berkeley National Laboratory, operated under Contract No. DE-AC02-05CH11231; the Texas Advanced Computing Center (TACC) at The University of Texas at Austin; the Extreme Science and Engineering Discovery Environment (XSEDE), which is supported by National Science Foundation Grant No. ACI-1548562; and Jefferson Lab.
\end{acknowledgments}

\vspace{0.5cm}

\noindent \textbf{Data Access Statement}

\vspace{0.2cm}

\noindent
Results of the fits to the $Q^2$ dependence of the form factors are given in the supplemental material. Reasonable requests for further data such as matrix elements or correlation functions can be directed to the authors and will be considered in accordance with the Hadron Spectrum Collaboration's policy on sharing data.


\bibliography{references}
\bibliographystyle{JHEP}

\appendix
\input{normalizing_two_points}
\input{results_appendix.tex}

\end{document}

%% file: title.tex
\title{Radiative Transitions in Charmonium from Lattice QCD}

\author[a]{James Delaney,} \emailAdd{jd688@cam.ac.uk}
\author[a]{Christopher~E.~Thomas,} \emailAdd{c.e.thomas@damtp.cam.ac.uk}
\author[b]{Sinéad M. Ryan} \emailAdd{ryan@maths.tcd.ie}
\author{\\(for the Hadron Spectrum Collaboration)}

\affiliation[a]{DAMTP, University of Cambridge, Centre for Mathematical Sciences, Wilberforce Road, Cambridge, CB3 0WA, UK}
\affiliation[b]{School of Mathematics and Hamilton Mathematics Institute, Trinity College, Dublin 2, Ireland}

\abstract{The coupling of various charmonium mesons to a photon is studied using lattice QCD, giving access to radiative form factors and transitions, and probing the mesons' structure. Methods are developed which allow the robust determination of amplitudes, including those involving an excited state as well as multiple form factors, for a range of kinematics. These are applied in dynamical lattice QCD calculations using the distillation technique to compute the underlying three-point correlation functions. Form factors and amplitudes involving the low-lying charmonia, $\eta_c$, $\Jpsi$, $\chi_{c0}$ and $\eta_c'$, are calculated, demonstrating the methods and providing the first dynamical lattice QCD results for some quantities.
}

\arxivnumber{2301.08213}

%% file: introduction.tex
\section{Introduction}
\label{sec:introduction}

The study of charmonium and charmonium-like mesons provides a useful window into the non-perturbative regime of QCD. The mass scale of the charm quark, falling between that of light quarks and the bottom quark, means that these systems are reasonably non-relativistic and hence cleaner to study than light hadrons theoretically, but they are light enough to be accessible in many different experiments.
A plethora of charmonia have been observed~\cite{Zyla:2020zbs}, including a number of states, known as exotic hadrons, that do not fit into the quark model picture of a meson as a bound state of a quark and an antiquark (see e.g.~\cite{Brambilla:2019esw, Chen:2016qju, Godfrey:2008nc} for some reviews).

Spectroscopy, the study of the masses and widths of stable hadrons and hadronic resonances, is one tool that can be used to probe QCD. However, more detailed information on the structure of hadrons can be obtained from other quantities; for example, from radiative transitions and electromagnetic form factors, related to a hadron's coupling to a photon.
Radiative partial widths such as $\Gamma(\Jpsi \to \eta_c \, \gamma)$ and $\Gamma(\chi_{c0} \to \Jpsi \, \gamma)$ have been measured by, and are continuing to be studied by, a range of experiments.
To complement experimental measurements, it is important to perform theoretical calculations to compare against and
aid in the interpretation of experimental results, e.g.\ to determine whether a state is a conventional charmonium meson, a hybrid meson, a molecular state, a tetraquark, etc. There is a long history of computing radiative transition amplitudes involving charmonia in phenomenological models, but first-principles QCD-based calculations are also necessary.
Lattice QCD, where the theory is discretised on a lattice in a finite spacetime volume and quantities are computed numerically, is a method that allows for systematically-improvable computations within QCD. Lattice calculations of a few radiative transition amplitudes involving low-lying charmonia, such as $\Jpsi \to \eta_c \, \gamma$, $\psi(2S) \to \eta_c \, \gamma$, $\chi_{c0} \to \Jpsi \, \gamma$, are now mature~\cite{Dudek:2006ej, Chen:2011kpa, Donald:2012ga, Becirevic:2012dc, Li:2018sfo, Li:2020gau, Li:2021gze}, but there have been fewer calculations involving higher-lying and excited charmonia~\cite{Dudek:2009kk, Yang:2012mya, Becirevic:2014rda}.

The aim of this paper is to establish and test a robust methodology that can be applied to radiative transitions involving exotic and excited charmonia, building on work on transitions in the light-quark sector~\cite{Shultz:2015pfa}, studies of the charmonium spectrum~\cite{HadronSpectrum:2012gic,Cheung:2016bym}, and earlier quenched calculations of radiative transitions involving charmonia~\cite{Dudek:2009kk,Dudek:2006ej}. The work in the charmonium sector showed how lattice calculations and their phenomenological interpretation can be used to probe the structure of charmonia, e.g.\ to identify a hybrid meson candidate. We will discuss how amplitudes and form factors can be robustly extract from computations of three-point correlation functions involving carefully constructed interpolating operators.
To demonstrate the techniques, transitions involving some low-lying charmonia, $\eta_c$ ($J^{PC}=0^{-+}$), $\Jpsi$ ($1^{--}$), $\chi_{c0}$ ($0^{++}$) and $\eta_c'$ ($0^{-+}$), are studied in dynamical lattice QCD calculations.

The rest of this paper is organised as follows.
In Section \ref{sec:RT_FF}, radiative transitions, form factors and the Lorentz structure of matrix elements are discussed. Section \ref{sec:method} presents the techniques employed to extract the relevant information from lattice QCD calculations and Section \ref{sec:calc_details} provides specific details of the calculations used in this study.
Results from applying the methodology are presented in Section \ref{sec:results} and a summary is given in Section \ref{sec:conclusion}.

%% file: transitions.tex
\section{Radiative Transitions and Form Factors}
\label{sec:RT_FF}

The coupling of hadrons to photons induces radiative transitions between hadrons. At leading order in QED, the amplitude for $h \to h' \gamma$ is proportional to the transition matrix element involving the electromagnetic current,
\begin{align}
    \bra{h'_{J'}(\lambda',\vec{p}^{\, \prime})} j^{\mu} \ket{h_{J}(\lambda, \vec{p} \, )} ,
\end{align}
where the hadrons $h, h'$ have angular momentum $J, J'$, helicity $\lambda, \lambda'$ and 4-momentum $p^{\mu}=(E_h,\vec{p}),\; p'^{\mu}=(E_h',\vec{p}^{\,\prime})$. The electromagnetic current is $j^{\mu} = \sum_i q_i \, \bar{\psi}_i \gamma^\mu \psi_i$ with quark charge $q_i$ and the sum taken over quark flavours $i$.
This matrix element can be decomposed as,
\begin{align}
    \label{eq:kin_decomp}
    \bra{h'_{J'}(\lambda',\vec{p}^{\, \prime})} j^{\mu} \ket{h_{J}(\lambda, \vec{p} \, )} = \sum_i K^{\mu}_i[h'_{J'}(\lambda',\vec{p}^{\, \prime}); h_{J}(\lambda, \vec{p} \, )] \, F_i(Q^2),
\end{align}
where $K^{\mu}_i[h'_{J'}(\lambda',\vec{p}^{\, \prime}); h_{J}(\lambda, \vec{p} \, )]$ are Lorentz vectors referred to as kinematic factors and $F_i(Q^2)$ are Lorentz scalars referred to as transition form factors (or just form factors if $h'=h$).
The form factors only depend on $Q^2 = -(E_h - E_h')^2 + \left|\vec{p} -\vec{p}^{\, \prime}\right|^2$, the virtuality of the photon.

The decomposition and kinematic factors for a particular matrix element can be constructed by considering all possible contractions of the polarization tensors, $\epsilon$, and 4-momenta associated with each hadron, provided that the contracted object carries a single Lorentz index, i.e.\ it has the correct Lorentz structure, and is linear in the polarization tensors.
Constraints from Ward identities and discrete symmetries such as parity must then be taken into account, removing form factors that are redundant or not allowed.

As an example, the matrix element of the pseudoscalar meson, $h'=h=\eta_c$, has a decomposition involving a single form factor,
\begin{align}
    \label{eq:etace_decomp}
    \bra{\eta_c(\vec{p}^{\, \prime})} j^{\mu} \ket{\eta_c(\vec{p}\,)} = (p + p')^\mu F(Q^2).
\end{align}
For a vector meson, $h'=h=\Jpsi$, there are three independent form factors $G_1, G_2, G_3$ such that
\begin{align}
    \bra{ J/\psi(\lambda',\vec{p}\,') } j^\mu \ket{ J/\psi (\lambda, \vec{p}) }
    = & - \big[(p+p')^\mu \; \epsilon^*(\lambda',\vec{p}\,') \cdot \epsilon(\lambda,\vec{p}\,) \big]\; G_1(Q^2) \nonumber                                                                               \\
      & + \big[\epsilon^\mu(\lambda,\vec{p}) \, \epsilon^*(\lambda',\vec{p}\,') \!\cdot\! p + \epsilon^{\mu*}(\lambda',\vec{p}\,')\, \epsilon(\lambda,\vec{p}\,)\!\cdot\! p' \big]\, G_2(Q^2) \nonumber \\
      & - \big[(p+p')^\mu \; \epsilon^*(\lambda',\vec{p}\,') \!\cdot\! p \;  \epsilon(\lambda,\vec{p}\,)\!\cdot\! p' \, \tfrac{1}{2m^2} \big]\; G_3(Q^2), \label{eq:psice_kindecomp}
\end{align}

There is no unique decomposition of a matrix element in terms of form factors: as well as the freedom to divide out a form factor by a Lorentz scalar and put it in the corresponding kinematic factor, when more one form factor is present any linearly independent set of form factors can be used.
In this paper we generally use the multipole basis \cite{Durand:1962zza}.

From the form factors we can infer qualitative and quantitative information on the hadrons. For example, by looking at the hierarchy of form factors partial radiative widths and charge distributions which are useful for comparison to experiment can be determined.
The partial radiative width for $h \to h' \gamma$ can be related to the multipole form factors by~\cite{Dudek:2009kk},
\begin{align}
    \label{eq:generic_partial_width}
    \Gamma(h \to h' \gamma) = \frac{1}{4\pi (2 J + 1)} \frac{|\vec{q}\,|}{m^{2}} \sum_k | F_k(0)|^2,
\end{align}
where $\vec{q} = \vec{p}^{\, \prime} - \vec{p}$ is the spatial component of the photon momentum, $m$ and $J$ are the mass and angular momentum of hadron $h$ respectively, and the sum is over all the relevant transverse multipole form factors, $\{ F_k(Q^2) \}$.
In the next section we describe how form factors are determined in lattice QCD calculations.

%% file: two_points.tex
\section{Methodology}
\label{sec:method}

Lattice QCD calculations are performed in a finite volume and so the spectrum is discrete.
If $h$ and $h'$ are stable with respect to QCD (i.e.\ they do not decay strongly), and $\estate{n}_i$ and $\estate{n}_f$ are the corresponding finite-volume energy eigenstates then matrix elements describing a transition $h \to h' \gamma$ are determined by considering three-point correlators of the form, 
\begin{align}
    \label{eq:generic_three_point_correlator}
    C^{\mu}_{\estate{n}_f \estate{n}_i}(\Delta t, t)
     & =  \bra{0} \Omega_{\estate{n}_f}(\Delta t) \, j^{\mu}(t) \, \Omega_{\estate{n}_i}^{\dagger}(0) \ket{0},
\end{align}
where $\Omega_{\estate{n}_i}^{\dagger}(0)$ and $\Omega_{\estate{n}_f}(\Delta t)$ are operators located on timeslices $0$ and $\Delta t$ which create $\estate{n}_i$ and annihilate $\estate{n}_f$ respectively.
The reduced symmetry of the finite-volume lattice compared to an infinite-volume continuum introduces some complications. Spin, $J$, is no longer a good quantum number and instead states are classified by the irreducible representations (irreps) of the relevant finite symmetry group. All the operators used in this work are projected onto definite momentum and constructed to transform in the appropriate irrep as described in \cite{Dudek:2010wm,Thomas:2011rh}; more details of the consequences of the reduced symmetry for extracting matrix elements can be found in Section D of \cite{Shultz:2015pfa}.
Before discussing how to extract matrix elements from the correlators, we will first describe the construction of operators
that efficiently interpolate the states of interest and this requires an analysis of two-point correlators.

\subsection{Lattice Spectroscopy and Optimized Operators}
\label{sec:2pts}
We form an `optimized operator' $\Omega_\estate{n}^\dagger$ as a linear combination of a large basis of operators $\{\mathcal{O}_i^\dagger\}$, $\Omega_\estate{n}^\dagger \sim \sum_i v_i \mathcal{O}_i^\dagger$,
where coefficients $\{v_i \}$ are found such that there is a large overlap $\bra{\estate{n}} \Omega_\estate{n}^\dagger \ket{0}$ onto the desired state $\estate{n}$, but very small overlap with other states.
Following the method described in \cite{Dudek:2010wm,Thomas:2011rh}, in each channel a large basis of
fermion-bilinear operators is used, built from Dirac gamma matrices and gauge-covariant derivatives, projected onto the desired quantum numbers and momentum and transforming in the appropriate finite-volume irrep.
With $N$ operators 
an $N \times N$ matrix of two-point correlators can be computed,
\begin{align}
    \label{eq:two_point_matrix}
    C_{ij}(t) = \bra{0} \mathcal{O}^{\,}_i(t) \, \mathcal{O}_j^\dagger(0) \ket{0}.
\end{align}
Optimal coefficients are then determined following the methods of \cite{Dudek:2007wv, Michael:1985ne}: a generalized eigenvalue problem for the correlator matrix is solved,
\begin{align}
    C_{ij}(t) \, v_j^{(\estate{n})} = \lambda_\estate{n}(t, t_0) \, C_{ij}(t_0) \, v_j^{(\estate{n})}, \label{eq:gevp}
\end{align}
where $t_0$ is an appropriate reference timeslice.
The generalized eigenvalues, $\lambda_\estate{n}(t, t_0)$, are related to the energies of the states as $\lambda_\estate{n}(t) \sim e^{-E_\estate{n}(t-t_0)}$, where $E_\estate{n}$ is the energy of the $\estate{n}$'th state in this channel.
The generalized eigenvectors $v_i^{(\estate{n})}$ give the variationally-optimal coefficients and optimised operators are then constructed as,
\begin{equation}
    \Omega_\estate{n}^\dagger = \sqrt{2E_\estate{n}} \, e^{-E_\estate{n}t_0/2} \sum\nolimits_i v^{(\estate{n})}_i \mathcal{O}_i^\dagger. \label{eq:opt_op_defn}
\end{equation}
The normalisation of the operator, i.e.~the overlap $\bra{\estate{n}} \Omega_\estate{n}^\dagger \ket{0}$, depends on the details of the implementation of the variational method and the choice of $t_0$ -- this is discussed in Appendix \ref{sec:overlap_factors}.

\begin{figure}[t]
    \centering
    \includegraphics[width=\textwidth]{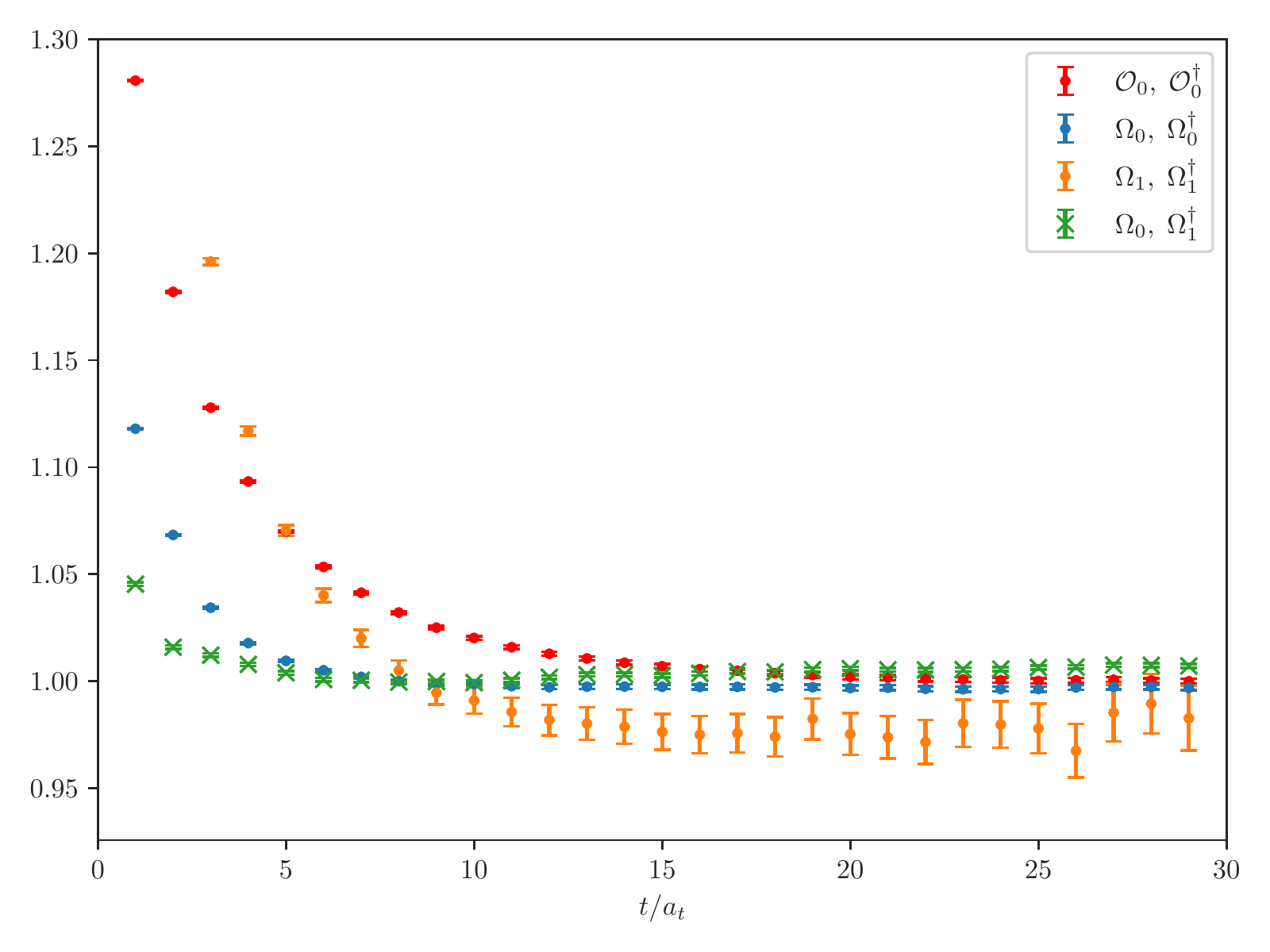}
    \caption{Illustration of the effectiveness of optimised operators in the $\eta_c$ channel at zero momentum.
        Points show correlators with the leading time dependence divided out as described in the text.
        Shown in blue and orange are correlators with optimised operators $\Omega_0$ and $\Omega_1$ designed to overlap onto the $\eta_c$ and $\eta_c'$ respectively at both source and sink.
        In red is the correlator obtained with a simple fermion bilinear operator $\mathcal{O}_0 \sim \bar{\psi} \gamma^5 \psi$ at both source and sink, normalised so that the plateau value is 1 at large time.
        In green is the ``off-diagonal'' correlator with $\Omega_1$ at the source and $\Omega_0$ at the sink -- these points have been shifted up by 1 to display them on the same plot.}
    \label{fig:proj_op_plot}
\end{figure}

As an example to demonstrate the effectiveness of these optimised operators, Figure~\ref{fig:proj_op_plot} shows two-point correlators in the $\eta_c$ channel at zero momentum with the leading time dependence divided out, i.e.~$C(t) \, e^{E_\estate{n}(t-t_0)}$, computed using the lattice setup described below in Section~\ref{sec:calc_details}.
The correlator with an optimised ground-state $\eta_c$ operator $\Omega_0$ at the source and sink (blue points) plateaus at a much earlier time than the correlator from a simple fermion-bilinear operator $\mathcal{O}_0 \sim \bar{\psi} \gamma^5 \psi$ (red points) with the plateau starting at $t/a_t \approx 10$ compared to $t/a_t \approx 18$.
This means that signals can be extracted at earlier times where the correlator is more statistically precise and for three-point correlators this enables the matrix element to be determined more precisely.
In addition, a good signal is found with the optimised operator $\Omega_1$ for the first excited state $\eta_c'$ (orange points), something which would not be possible with a simple fermion-bilinear operator.
This plateau value can be slightly less than 1 because of the normalisation of the operators -- for more details see the discussion around Eq.~\eqref{eq:two_point_norm_factor} in Appendix \ref{sec:overlap_factors}.
The green points show the correlator with $\Omega_1^\dagger$ at the source and $\Omega_0$ at the sink -- as well as having the leading time dependence divided out, these have been shifted up by 1 for display on the same plot. This early plateau to 1, i.e.\ plateau of unshifted correlator to 0,
demonstrates the good orthogonality between these two operators.

%% file: three_points.tex
\subsection{Correlator Analysis and Extracting Form Factors}
\label{sec:plat_fits}
\label{sec:multi_kin}
Having established how to construct operators that efficiently interpolate states of interest, we now turn focus back to three-point correlators of the form in eq.~\eqref{eq:generic_three_point_correlator}.
Exposing the overlap of the operator onto the desired state and a single additional state, \mbox{$\Omega_{\estate{n}_i}^\dagger \ket{0} \sim \ket{\estate{n}_i} + \varepsilon_{\estate{n}'_i} \ket{\estate{n}'_i} + \dots$}, where $\varepsilon_{\estate{n}'_i}$ is small and the ellipsis denotes other suppressed terms from overlaps onto higher states, we can expand the correlator as
\begin{align*}
    C^{\mu}_{\estate{n}_f \estate{n}_i}(\Delta t, t) = & \; e^{-E_{\estate{n}_f}(\Delta t - t)} e^{-E_{\estate{n}_i}t} \bra{\estate{n}_f} j^{\mu}(0) \ket{\estate{n}_i} + \varepsilon_{\estate{n}'_f}^* e^{-E_{\estate{n}'_f}(\Delta t - t)} e^{-E_{\estate{n}_i}t} \bra{\estate{n}'_f} j^{\mu}(0) \ket{\estate{n}_i} \\
                                                       & + \varepsilon_{\estate{n}'_i} e^{-E_{\estate{n}_f}(\Delta t - t)} e^{-E_{\estate{n}'_i}t} \bra{\estate{n}_f} j^{\mu}(0) \ket{\estate{n}'_i} + \dots \, .
\end{align*}
The desired matrix element can then be extracted from
\begin{align*}
    \Gamma(Q^2; t) = e^{E_{\estate{n}_f}(\Delta t - t)} \, e^{E_{\estate{n}_i}t} \, C^{\mu}_{\estate{n}_f \estate{n}_i} = \bra{\estate{n}_f} j^{\mu}(0) \ket{\estate{n}_i}
    + f_{\estate{n}'_f} \, e^{-\delta E_f \, (\Delta t - t)}
    + f_{\estate{n}'_i} \, e^{-\delta E_i \,t} + \dots \, ,
\end{align*}
where terms involving excited state matrix elements are combined in the $f_{\estate{n}'_f}, f_{\estate{n}'_i}$ terms.
It can be seen that at appropriate intermediate times
$ 0 < t < \Delta t$, the second and third terms should be highly suppressed and the correlator
should plateau to a value corresponding to the matrix element.

\begin{figure}[t]
    \centering
    \includegraphics[width=\textwidth]{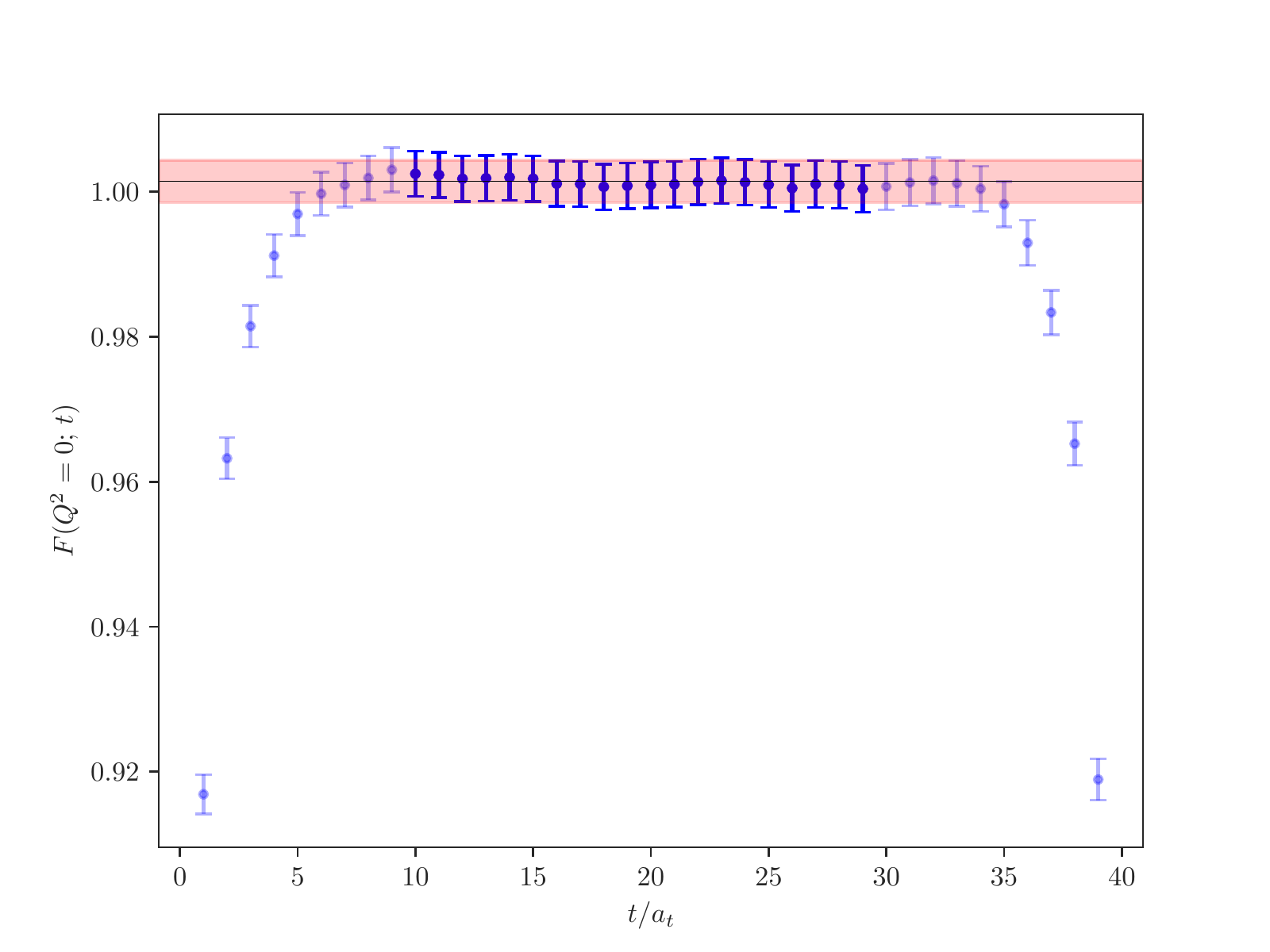}
    \caption{ $F(Q^2=0,t)$ obtained from a three-point correlator featuring an optimised $\eta_c$ operator with $\vec{p} = \tfrac{2 \pi}{L}(0,0,1)$ at both the source and the sink. The black line shows the result of a fit to a constant for $10 \le t/a_t \le 30$ with the band indicating the $\pm 1 \sigma$ statistical uncertainty (darker points are included in fit whilst translucent points are not). }
    \label{fig:sample_correlator}
\end{figure}

The general kinematic decomposition outlined in eq.~(\ref{eq:kin_decomp}) shows that multiple form factors can contribute to a single matrix element.
If only a single kinematic factor appears,
then dividing the matrix element by the kinematic factor, $K^\mu$, for $K^\mu \neq 0$, yields the desired form factor,
\begin{align*}
    F(Q^2) = \frac{1}{K^\mu}  \bra{\estate{n}_f} j^{\mu}(0) \ket{\estate{n}_i} ,
\end{align*}
where there is no summation over $\mu$.
In this case, we consider a time-dependent form factor, $F(Q^2; t) = \Gamma(Q^2; t) / K^\mu$, and fits of the time dependence to a constant, $F(Q^2; t)=F(Q^2)$, i.e.\ assuming contamination is negligible, and to
\begin{equation*}
    F(Q^2; t) = \left\{ \begin{array}{ll}
        F(Q^2) + f_n e^{-\delta E_n(\Delta t - t)} + f_n e^{- \delta E_n t} \quad\quad & \estate{n}_i = \estate{n}_f \\
        F(Q^2) + f_n e^{-\delta E_n(\Delta t - t)} + f_m e^{- \delta E_m t} \quad\quad & \text{otherwise} \, .
    \end{array}
    \right.
\end{equation*}
The final fit form and time range used are chosen to achieve a good fit while permitting as large a time range as possible.
When defining the $\chi^2$ to be minimised in the fitting procedure, the covariance matrices used to account for correlations between different timeslices are estimated using a ``shrinkage'' approach presented in \cite{stein1956variate, Chen:2010, Rinaldi:2019thf, Ledoit:2000}.
In particular, we use an implementation \cite{scikit-learn} of the ``Oracle Approximating Shrinkage Estimator'' for the covariance matrix.\footnote{There are various methods proposed in the literature to obtain a better-conditioned estimate of the covariance matrix, each with advantages and disadvantages. Different approaches were considered and the chosen one has some positive features, tends to the true covariance matrix in the limit of infinite statistics and does not require choosing an ad-hoc parameter.}

As an example, Figure~\ref{fig:sample_correlator} shows $F(Q^2=0,t)$ obtained from a three-point correlator featuring an optimised $\eta_c$ operator with $\vec{p} = \tfrac{2 \pi}{L}(0,0,1)$ at both the source and the sink, computed using the lattice setup described below in Section~\ref{sec:calc_details}, where $L$ is the spatial extent of the lattice. This can be seen to plateau rapidly at intermediate times, $0 < t < \Delta t$, allowing a fit over a large time range and so enabling the form factor to be extracted robustly and with high statistical precision.

To allow propagation of correlations into subsequent analyses, such as the $Q^2$ dependence of the form factor, the fitting procedure is repeated on jackknife samples of the data.
This allows an additional verification of the robustness of the fit under small perturbations. 
Fits to subsets of data consisting of only correlators containing a temporal current or only those containing a spatial current are also considered.

When multiple form factors enter, there is in general an under-specified linear system at each $Q^2$ because a single matrix element cannot determine multiple form factors.
In this case we bin together sets of matrix elements with the same or similar values of $Q^2$.
There is a compromise between the number of data points in a given bin and the range of $Q^2$ that a bin spans.
Bigger bins provide more constraints on the form factors in a bin, but at the expense of assuming the form factors do not vary appreciably over a larger $Q^2$ interval (i.e.\ over the bin) and having fewer form factor estimates across the range of $Q^2$.

From the relativistic dispersion relation,
\begin{align*}
    Q^2 = -q^2 = -\left(\sqrt{m_i^2 + \vec{p}^{\, 2}} - \sqrt{m_f^2 + \vec{p}^{\, \prime \, 2}} \right)^2 + (\vec{p} - \pprime)^2,
\end{align*}
it can be seen that if two matrix elements have the same initial momentum and the same final momentum they will have the same $Q^2$.
Such a redundancy can be generated by varying the Lorentz index of the current and the spin/helicity component of any hadrons with non-zero spin.
A more coarse-grained approach is to group matrix elements that have the same magnitude of current momentum $\vec{q} = \vec{p} - \pprime$ but different $\vec{p}$, $\pprime$.
If $m_i \approx m_f$ and $m_i^2 \gg \vec{p}^{\, 2}$, $\vec{p}^{\, \prime \, 2}$ then these matrix elements will have similar $Q^2$.

Enumerating the matrix elements within a $Q^2$ bin with an index $a$, the linear system can be written as,
\begin{align*}
    \Gamma_a (Q^2; t) = \sum_i K_{ai} \, F_i (Q^2; t), 
\end{align*}
where $i$ enumerates the form factors in the decomposition.
Expressing the matrix of kinematic factors as $\mathsf{K}$ and the vectors of form factors and matrix elements as $\mathsf{F}$ and $\mathsf{\Gamma}$, these form factors can be determined by linear regression \cite{Shultz:2015pfa} using the normal equation,
\begin{align}
    \label{eq:norm_equation}
    \mathsf{F} (Q^2; t) = (\mathsf{K}^\dagger \, \mathsf{K})^{-1} \, \mathsf{K}^\dagger \, \mathsf{\Gamma} (Q^2; t).
\end{align}
We then fit the time dependence of the form factor to obtain $F_i(Q^2)$ as in the case described above where only a single form factor appears.
To check the robustness of the form factor determinations, we repeat the process a number of times systematically removing different matrix elements in the bin from consideration.
Equation \eqref{eq:norm_equation} is solved on each time slice and each gauge configuration (via jackknife) to yield $N_{\textrm{cfgs}}$ samples of $F_i(Q^2; t)$.
This binning procedure can also be used when only a single form factor appears, but in this case it is generally preferable to consider each matrix element separately because this allows more straightforward independent cross-checks, and the isolation of contributions from different lattice irreps and from the spatial current v.s.\ the temporal current.

\begin{table}[t]
    \centering
    \renewcommand{\arraystretch}{1.8}
    \begin{tabular}{l|l}
        Fit form label & $F(Q^2; ...)$                                                      \\
        \hline
        Exp1           & $F_0 \, e^{-\frac{Q^2}{16 \beta^2}}$                              \\
        Exp2           & $F_0 \left(1 + \alpha Q^2 \right) e^{-\frac{Q^2}{16 \beta^2}}$    \\
        Exp3           & $F_0 \, e^{-\frac{Q^2}{16 \beta^2} \left(1 - \alpha Q^2 \right)}$ \\
        VMD1           & $F_0 \, \frac{1}{1 + Q^2 / m^2}$                                  \\
        VMD2           & $F_0 \, \frac{1}{1 + Q^2 / m^2 + \alpha (Q^2 / m^2)^2 }$          \\
        ExpVMD         & $F_0 \, \frac{1}{1 + Q^2 / m^2} \, e^{-\frac{Q^2}{16 \beta^2}} $  \\
    \end{tabular}
    \caption{Forms considered when fitting the $Q^2$ dependence of form factors in this study.}
    \label{tab:fit_forms}
\end{table}

With multiple form factors an alternative approach would be to fit the time dependence of $\Gamma_a (Q^2; t)$ to get $\Gamma_a (Q^2)$, and from this obtain $F_i(Q^2)$. A potential advantage of this method
is that the expected time dependence of contamination from unwanted states in $\Gamma_a (Q^2; t)$ is clearer
than in $F_i(Q^2; t)$. However, because we use optimised operators, we find that matrix elements plateau a relatively short distance away from the source/sink (see Fig.~\ref{fig:proj_op_plot}) and a reliable plateau in $F_i(Q^2; t)$ is consistently found. We find that our procedure leads to more stable determinations of the form factors -- the alternative method requires very many fits to the time dependence and when a form factor is associated with a small kinematic factor a small change to one fit can have a significant impact on the result. Where both methods give robust determinations of the form factors, the results from each are consistent.

Equipped with a set of form factors $\{ F(Q^2) \}$, the $Q^2$ dependence can be parameterised to enable extrapolation/interpolation to $Q^2=0$.
To avoid model dependence from assuming a particular $Q^2$ form, we attempt fits with the variety of phenomenologically-motivated fit forms listed in Table~\ref{tab:fit_forms}. As discussed later, we discard any fits that are significantly worse at describing the data.
Correlations are again taken into account through a covariance matrix that is estimated using the shrinkage approach as described above.

%% file: calc_details.tex
\section{Calculation Details}
\label{sec:calc_details}

\begin{table}[t]
    \centering
    \begin{tabular}{|l|l|l|l|l|l|}
        \hline
        $(L/a_s)^3 \times (T / a_t)$ & $m_{\pi}$ / MeV & $N_{\textrm{cfgs}}$ & $N_{\textrm{vecs}}$ & $a_s$ / fm & $m_\pi L$ \\ \hline
        $20^3 \times 128$            & 391             & 603                 & 128                 & 0.12       & 4.8       \\ \hline
    \end{tabular}
    \caption{Lattice parameters for the lattice used in this study. $N_{\text{vecs}}$ is the number of distillation vectors.}
    \label{tab:lat_params}
\end{table}

This study uses a single ensemble of gauge fields, generated with a tree-level Symanzik-improved gauge action and a tadpole-improved Sheikholeslami-Wohlert (clover) fermion action with stout smeared spatial gauge fields and 2+1 flavours of dynamical quarks (two degenerate light up and down quarks, and a heavier strange quark)~\cite{Edwards:2008ja, Lin:2008pr}.
Correlation functions were computed on 603 gauge configurations.
The discretisation is anisotropic with a spatial lattice spacing, $a_s \sim 0.12 \, \mathrm{fm}$, and a finer temporal lattice spacing, $a_t$, with anisotropy $\xi = a_s/a_t \sim 3.5$.
The lattice volume is $(L/a_s)^3 \times (T/a_t) = 20^3 \times 128$ -- this relatively large volume with $m_\pi L \sim 5$ means that finite-volume effects should not be significant for charmonia below $D\bar{D}$ threshold as can be seen in \cite{HadronSpectrum:2012gic}.
The cubic spatial volume with periodic boundary conditions quantizes momentum as $\vec{p} = \frac{2 \pi}{L} (n_x, n_y, n_z)$, $n_i\in \mathbb{Z}$, and in this work we will consider momenta with $n^2 = n_x^2 + n_y^2 + n_z^2 \leq 4$.

\begin{figure}[t]
    \centering
    \begin{subfigure}[b]{0.22\textwidth}
        \centering
        \includegraphics[width=\textwidth]{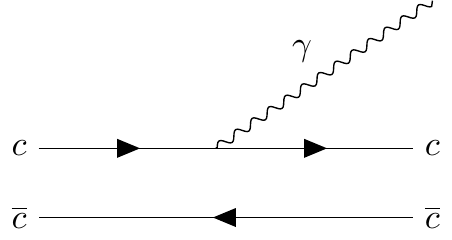}
        \caption{}
        \label{fig:feyn_diag_1}
    \end{subfigure}
    \hfill
    \begin{subfigure}[b]{0.22\textwidth}
        \centering
        \includegraphics[width=\textwidth]{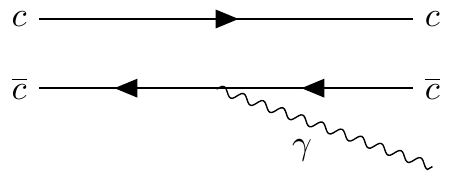}
        \caption{}
        \label{fig:feyn_diag_2}
    \end{subfigure}
    \hfill
    \begin{subfigure}[b]{0.22\textwidth}
        \centering
        \includegraphics[width=\textwidth]{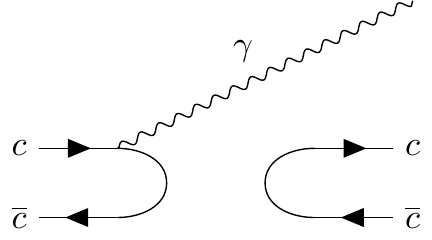}
        \caption{}
        \label{fig:feyn_diag_3}
    \end{subfigure}
    \hfill
    \begin{subfigure}[b]{0.22\textwidth}
        \centering
        \includegraphics[width=\textwidth]{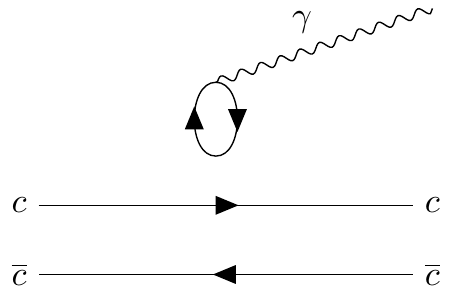}
        \caption{}
        \label{fig:feyn_diag_4}
    \end{subfigure}
    \caption{Relevant Wick contraction topologies. Diagrams (\ref{sub@fig:feyn_diag_1}) and (\ref{sub@fig:feyn_diag_2}) show fully connected diagrams, with the photon coupling to the charm quark and charm antiquark respectively. Diagrams (\ref{sub@fig:feyn_diag_3}) and (\ref{sub@fig:feyn_diag_4}) are disconnected. There is also another disconnected diagram like (\ref{sub@fig:feyn_diag_3}) where the photon couples to the sink rather than the source and another like (\ref{sub@fig:feyn_diag_4}) where the charm quark and antiquark annihilate as in (\ref{sub@fig:feyn_diag_3}). As discussed in the text, these disconnected diagram are expected to give only very small contributions and in this study only diagram (\ref{sub@fig:feyn_diag_1}) is computed. }
    \label{fig:feyn_diags}
\end{figure}

Strange quarks are tuned to approximate their physical mass while the unphysically-heavy light quarks yield $m_\pi \approx 391 \, \mathrm{MeV}$.\footnote{Charmonia below $D\bar{D}$ threshold are not expected to be very sensitive to the light-quark mass as observed in \cite{Cheung:2016bym}.}
The tuning of the parameters in the charm-quark action and the impact of any imperfect tuning are described in Ref.~\cite{HadronSpectrum:2012gic}.
To present results in physical units, the lattice spacing is determined using the mass of the $\Omega$ baryon which gives $a_t^{-1} = \frac{m_\Omega^{phys.}}{a_t m^{lat.}_\Omega} \, = \, 5.666 \, \mathrm{GeV}$ \cite{Edwards:2012fx}.
A summary of the lattice parameters can be found in Table \ref{tab:lat_params}.

We compute correlators using the distillation framework \cite{Peardon:2009gh} and further details on the calculation of three-point correlators in this approach can be found in \cite{Shultz:2015pfa}.
Quark fields at the source and the sink are smeared using the distillation operator
which projects onto the low energy modes.
The quark fields in the vector current insertion are unsmeared because otherwise a complicated energy-dependent renormalisation would be required to connect with physical amplitudes. The operators at the source and sink, and the current are each projected onto definite momentum, and the source-sink separation is $\Delta t / a_t = 40$.

Diagrams representing the relevant Wick contractions for the coupling of a photon to a meson are shown in Fig \ref{fig:feyn_diags}. In this work we only compute the contribution where the photon couples to the charm \emph{quark} (a) and not where the photon couples to the charm \emph{antiquark} (b), and we do not explicitly include the electric charge of the quark -- this means that we actually compute scaled versions of the multipole form factors, $\hat{F}_k$, defined by $F_k(Q^2) = \frac{4e}{3} \hat{F}_k(Q^2)$. It also implies that charge-conjugation ($C$) violating processes can be investigated, including the form factors of charmonia, and used to probe the structure of states.\footnote{Since charmonia are eigenstates of $C$, the photon couples equally to the charm quark and the charm antiquark, and the photon has $C=-1$, physically only radiative transitions between charmonia of opposite $C$ can occur.}
In addition, disconnected diagrams like (c) where the charm quark and antiquark annihilate are not computed -- these are OZI suppressed and are expected to only give small contributions in charmonium~\cite{Levkova:2010ft}.
We also neglect diagrams where the photon couples to a disconnected quark loop (d) which are again expected to only give small contributions -- the contribution from a charm quark loop is suppressed by the large charm-quark mass and in the SU(3) flavour symmetry limit the contributions from light and strange quark loops cancel.

Working with a finite lattice spacing leads to discretisation effects and with only one lattice spacing we can not fully quantify them.
The hyperfine splitting, $\Delta m = m_{\Jpsi} - m_{\eta_c}$, is particularly sensitive to these and on this lattice we find $\Delta m_\textrm{lat.} = 80.36(16) \mathrm{MeV}$ compared with the experimental value $\Delta m_\textrm{exp.} = 113.0(4) \mathrm{MeV}$ -- Ref.~\cite{HadronSpectrum:2012gic} discusses this tension and a rough estimate of the size of the discretisation uncertainties.

\begin{figure}~
    \centering
    \includegraphics[width=\textwidth]{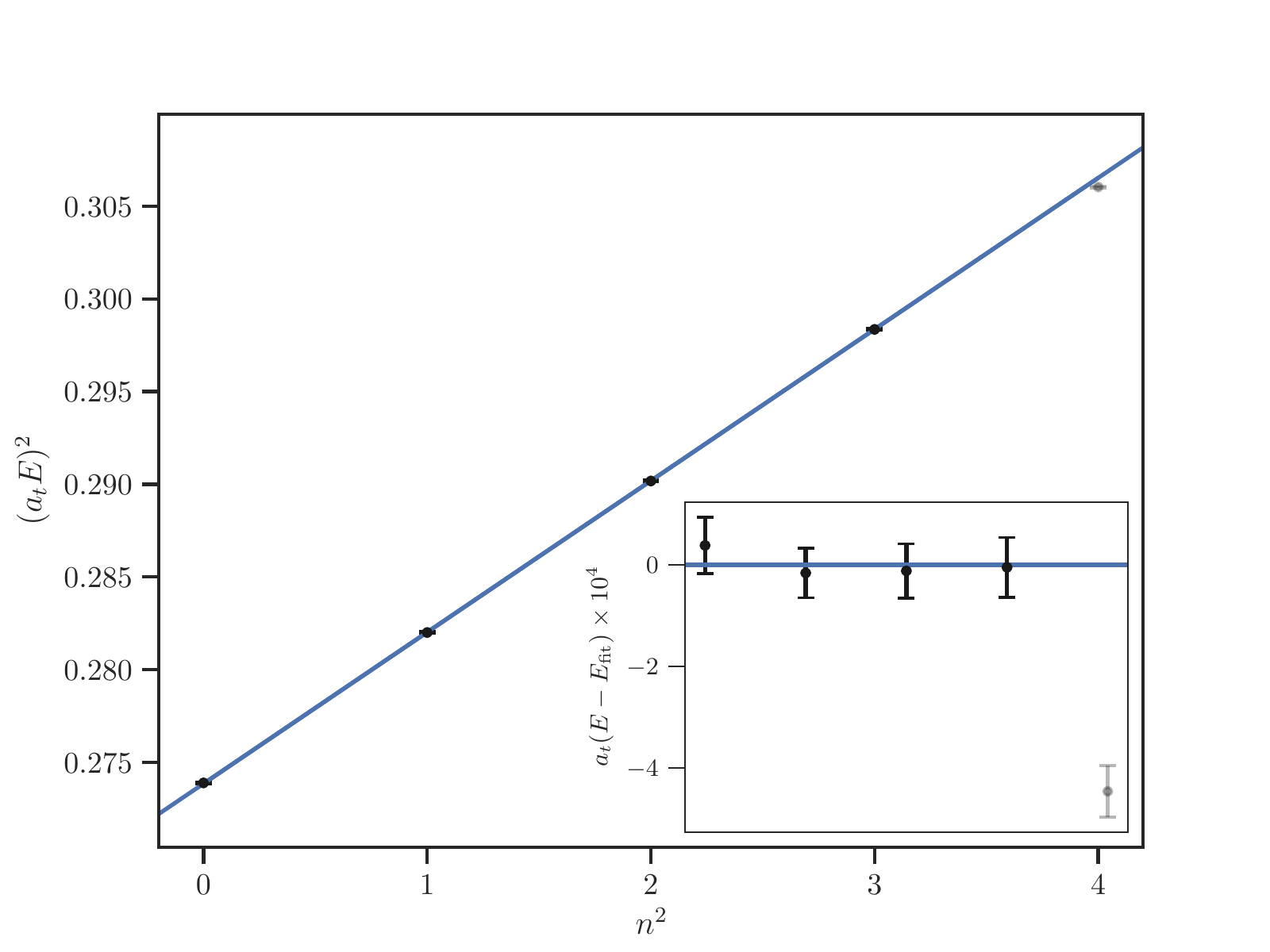}
    \caption{Momentum dependence of the $\eta_c$ energy on the lattice used in this study.
        The points show the computed energies and the line shows the result of a fit to the relativistic dispersion relation, eq.~\eqref{eq:dispersion_relation}, which has $\chi^2 / \Ndf = 1.64 / (4-2) = 0.82$.
        The point in grey with $n^2=4$ was not included in the fit.
        The inset shows the residuals, $a_t(E - E_\textrm{fit}) \times 10^4$. }
    \label{fig:etace_disp_20_no_4}
\end{figure}

We use an anisotropic lattice with different temporal and spatial lattice spacings and, while parameters in the action are tuned to restore hypercubic symmetry as far as possible, this leads to an additional type of discretisation effect. The anisotropy experienced by a stable hadron can be determined from the relativistic dispersion relation,
\begin{align}
    \label{eq:dispersion_relation}
    \left(a_t E\right)^2 =  \left(a_t m\right)^2 + \left(\tfrac{2\pi}{\xi L / a_s}\right)^2 n^2 ,
\end{align}
where $m$ is the mass of the hadron and $E$ is its energy at momentum magnitude $|\vec{p}\,| = \frac{2 \pi}{L} |\vec{n}|$.
The mass and anisotropy determined from fits to eq.~\eqref{eq:dispersion_relation} with $n^2 \leq 3$ for relevant charmonia are presented in Table \ref{tab:disp_relations} and, as an example, the momentum dependence of the $\eta_c$ energy is shown in Figure \ref{fig:etace_disp_20_no_4}.
The fits are generally found to describe the data reasonably well without the need to add higher-order discretization terms such as $(a_s p)^4$.
The measured anisotropies are all similar and in line with what has been found in other flavour sectors on this lattice~\cite{Ryan:2020iog, Moir:2013ub, HadronSpectrum:2012gic, Dudek:2012gj, Woss:2019hse}, suggesting that these discretisation effects are small.
The small discrepancy between the values of $\xi$ measured for the different helicity magnitudes of the $J/\psi$ ($\lambda = 0$ and $|\lambda| = 1$) is due to the reduced symmetry of the lattice and is consistent with what has been observed in other studies~\cite{Thomas:2011rh,Woss:2019hse,Woss:2018irj}.
To estimate the systematic effect from the small discrepancies between different determinations of the anisotropy, tests were performed with $\xi$ varied over an envelope of $\xi_\estate{n_i} \pm \sigma$, $\xi_\estate{n_f} \pm \sigma$, where $\sigma$ is the corresponding statistical uncertainty.
In all tests, the resulting effect is much smaller than the statistical uncertainty associated to the form factors.\footnote{We estimated the effect from the uncertainties on $m_\estate{n_i}$ and $m_\estate{n_f}$ in the same way, and reached the same conclusion.}
In addition, we avoid using higher momenta where discretisation effects are expected to become more important.
In the next section we discuss the renormalisation of the vector current and how we use an improved current to reduce discretisation uncertainties in calculations of the three-point correlation functions.

\begin{table}[hbp]
    \vspace{1cm}
    \centering
    \begin{tabular}{c|c|c|c}
        Hadron                  & $a_t m$     & $\xi$      & $\chi^2/\Ndf$ \\
        \hline
        $\eta_c$                & 0.52331(5)  & 3.4758(34) & 0.8           \\
        $J/\psi(\lambda = 0)$   & 0.53752(7)  & 3.4835(37) & 0.7           \\
        $J/\psi(|\lambda| = 1)$ & 0.53753(7)  & 3.4726(35) & 0.7           \\
        $\chi_{c0}$             & 0.60481(25) & 3.438(17)  & 0.14          \\
        $\eta'_c$               & 0.6437(7)   & 3.42(5)    & 1.4           \\
    \end{tabular}
    \caption{Relevant hadron masses and anisotropies from fits to eq.~\eqref{eq:dispersion_relation} for $n^2 \leq 3$.\protect\footnotemark}
    \label{tab:disp_relations}
\end{table}
\footnotetext{For the $J/\psi$ with $|\lambda| = 1$, there is tension between the $n^2 = 2$ energy in the $B_1$ irrep and the energy in the $B_2$ irrep, and so we use the mean of these two energies in the fit.}

%% file: misc_results.tex
\subsection{Renormalizing and Improving the Vector Current}
\label{sec:Zfactor}
\begin{figure}
  \centering
  \includegraphics[width=\textwidth]{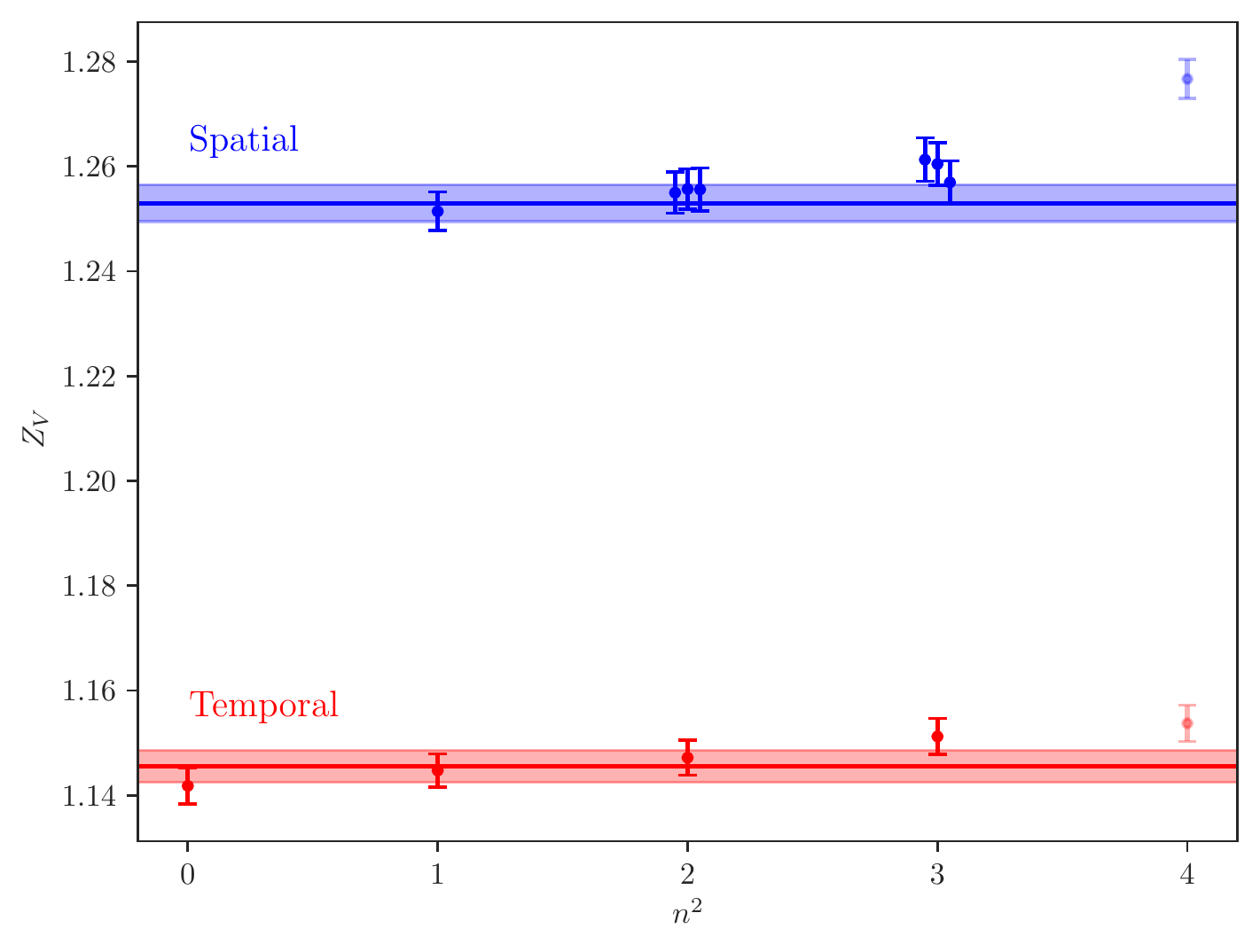}
  \caption{Renormalization factors as defined in eq.~\eqref{eq:ZV_defn} for the temporal current $Z_V^t$ (red points) and the spatial current $Z_V^s$ (blue points), for a range of equal source and sink momenta $\vec{p} = \left(\tfrac{2 \pi}{L}\right) \vec{n}$, determined using the methods discussed in Section \ref{sec:plat_fits}.
    Where multiple points occur at a given $n^2$ they are slightly horizontally displaced to aid visibility.
    The red and blue bands are from fits to a constant and show the mean and $\pm 1 \sigma$ statistical uncertainty.
    Points corresponding to correlators where the source and sink have $|\vec{p}\,|^2 \geq 4 \left(\tfrac{2 \pi}{L}\right)^2$ are translucent and are not used in the fits.}
  \label{fig:Z_factor}
\end{figure}

The local vector current
is not conserved in our lattice formulation and must be renormalised in order to relate the extracted matrix elements to physical results.
We perform a non-perturbative renormalization by imposing that $F_{\eta_c}(Q^2=0)$ be unity. The matrix elements we extract must therefore be multiplied by,
\begin{align}
  \label{eq:ZV_defn}
  Z_V = \frac{F^\textrm{cont.}_{\eta_c}(0)}{F^\textrm{lat.}_{\eta_c}(0)} = \frac{1}{F^\textrm{lat.}_{\eta_c}(0)} \, .
\end{align}
On an anisotropic lattice, the renormalisation constants for the temporal and spatial currents, $Z_V^t$ and $Z_V^s$, are distinct and must be determined seperately. Following the methods of Section \ref{sec:plat_fits}, $F^\textrm{lat.}_{\eta_c}(0)$ is extracted from $\bra{0} \Omega_{\eta_c}(\Delta t, \vec{p}) \, j^\mu (t, \vec{q}=0) \, \Omega^\dagger_{\eta_c}(0,\vec{p}) \ket{0}$ for a range of momenta $\vec{p}$ and the resulting $Z_V$ are shown in Figure \ref{fig:Z_factor}.
Fitting the data for $|\vec{p}\,|^2 \leq 3 \left(\tfrac{2 \pi}{L}\right)^2$ to a constant yields,
\begin{align*}
  Z_V^t = 1.145(3), \quad Z_V^s = 1.253(3),
\end{align*}
where $\chi^2/\Ndf = 14/(4-1) = 4.7$ and $\chi^2/\Ndf = 24/(7-1) = 4.0$ for the temporal and spatial fits respectively, and the uncertainties are statistical.
Results presented in subsequent sections have all been appropriately renormalized.

To reduce discretisation effects we add an $\mathcal{O}(a)$ improvement term to the local vector current which, for the fermionic action used here, leads to improved renormalised temporal and spatial vector currents \cite{Shultz:2015pfa},
\begin{align}
  j_0 & = Z^t_V \left( \bar{\psi}\gamma_0\psi + \frac{1}{4}\frac{\nu_s}{\xi}(1-\xi)a_s\partial_j(\bar{\psi}\sigma_{0j}\psi)\right), \nonumber \\
  j_k & = Z^s_V\left( \bar{\psi}\gamma_k\psi + \frac{1}{4}(1-\xi)a_t\partial_0(\bar{\psi}\sigma_{0k}\psi)\right),
  \label{eq:imp_current}
\end{align}
where $\nu_s=1.078$ is a parameter in the fermion action \cite{Edwards:2008ja} related to the spatial clover coefficient.
Note that the improvement term vanishes for $Q^2 = 0$ and so it does not modify the determination of $Z_V$.

\begin{figure}
  \centering
  \begin{center}
    \centerline{\includegraphics[width=1.2\textwidth]{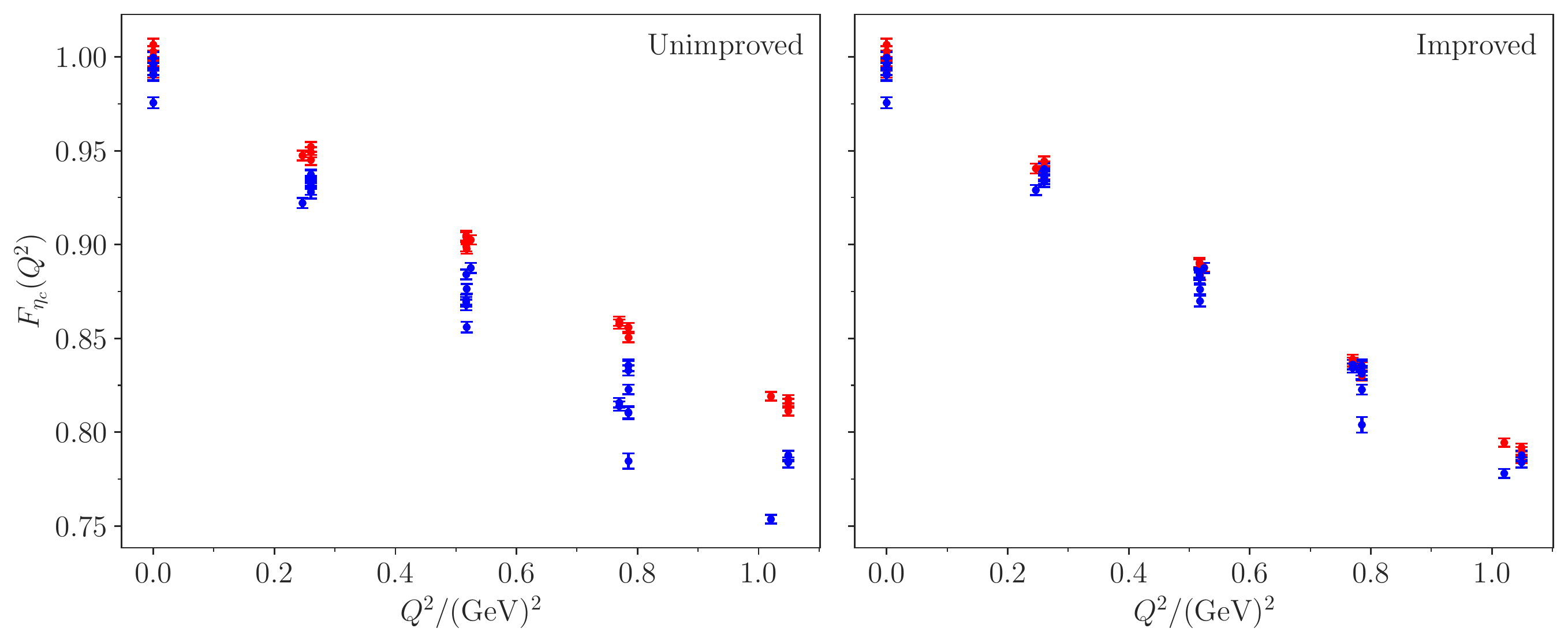}}
    \caption{The $\eta_c$ radiative form factor, $F_{\eta_c}(Q^2)$, defined in eq.~\eqref{eq:etace_decomp}, plotted against $Q^2$ calculated using the unimproved current (left plot) and the improved current (right plot). Each point corresponds to a single correlator involving a temporal (red) or spatial (blue) current with the methods of Section \ref{sec:plat_fits} used to extract the form factor and the appropriate renormalisation factor ($Z_V^t$ or $Z_V^s$) applied.
      The effectiveness of the improvement term can clearly be seen in the reduced spread of the form factors for each $Q^2$ in the right plot compared to the left plot.}
    \label{fig:improved_scatter_comparison}
  \end{center}
\end{figure}

The effect of the improvement term was investigated and it was found to improve the agreement between the
determination of physical observables from the spatial and temporal currents.
As a demonstration, the $Q^2$ dependence of the $\eta_c$ radiative form factor is shown using both the unimproved and the improved vector currents in Figure \ref{fig:improved_scatter_comparison}.
The most pronounced effect of the improvement is to reduce the discrepancy between results obtained from the temporal and spatial current insertions.
In addition, the spread for a given temporal/spatial current with different source and sink momentum combinations is reduced.
This effect appears to be more pronounced at larger $Q^2$ where the coefficient of the improvement term and discretization effects increase.
All the results we present in subsequent sections were obtained using the improved currents.

%% file: etace_etace.tex
\section{Results}
\label{sec:results}
In this section results for radiative and transition form factors involving some $S$ and $P$-wave charmonia are presented. In all cases the improved vector current was used and the renormalisation factor has been applied as discussed in Section \ref{sec:Zfactor}.
Comparisons to previous lattice calculations and experimental results are discussed where appropriate.
Our aim is to develop and test techniques that can be used to study transitions between various charmonia and to gain insight into their structure, and not to quantify all the systematic uncertainties.
The uncertainties quoted in final results are obtained by taking a range over the allowed fit forms, variations over the masses and anisotropies, and the statistical uncertainties.

\subsection{\texorpdfstring{$\eta_c$}{η} Form Factor}
\label{sec:etace_ff}

The simplest application of the methodology outlined above is to cases with a single form factor, and
perhaps the most straightforward to consider is the form factor of the lowest-lying charmonium meson, the $\eta_c$. The relevant kinematic decomposition is given in eq.~\eqref{eq:etace_decomp}.
This transition can not occur in nature since it violates charge-conjugation parity conservation, or, looked at another way, coupling of the current to the charm quark would cancel with the contribution from coupling to the anti-charm quark. Nevertheless, it can be determined in a lattice calculation by only coupling the current to the charm quark as discussed in Section~\ref{sec:calc_details}, serves as a useful test of lattice methods and can provide information about the meson.

\begin{figure}[tbp]
  \centering
  \includegraphics[width=\textwidth]{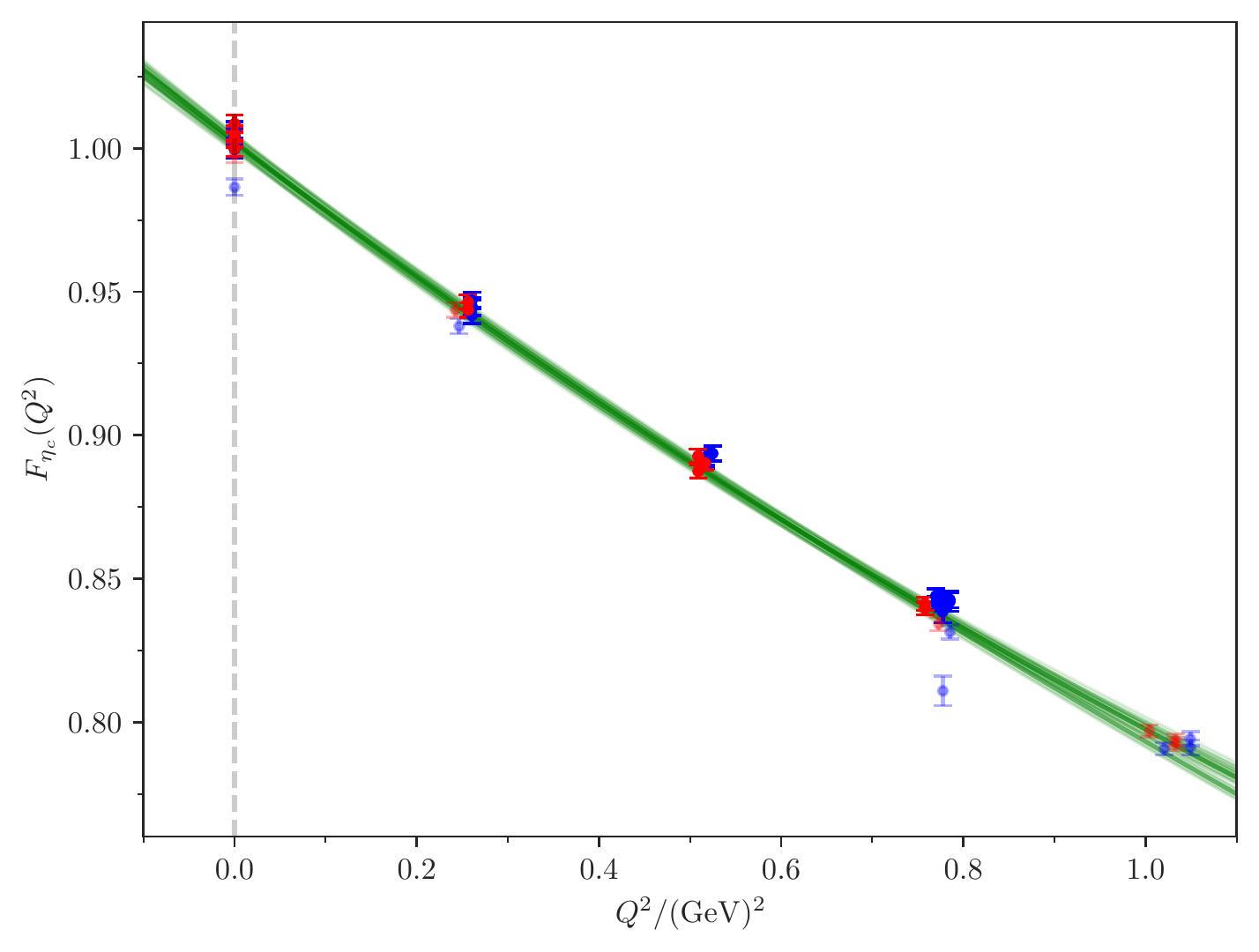}
  \caption{The $\eta_c$ form factor, $F_{\eta_c}(Q^2)$. Red and blue points are the form factor extracted from correlation functions featuring a temporal or spatial current insertion respectively. Points corresponding to correlators where at least one of the source, sink and current have momentum magnitude $\geq 2 \tfrac{2 \pi}{L}$ are translucent and not included in the fits.
    Bands show results of fits to the $Q^2$ dependence, as described in the text, giving $1\sigma$ statistical uncertainties around the mean.
    The dashed grey line indicates $Q^2 = 0$.}
  \label{fig:etace}
\end{figure}

To determine the form factor, three-point correlators were computed for a range of $Q^2$ values by varying the source and sink momenta, $\vec{p}$ and $\pprime$.
Following the approach outlined above, each correlator was divided by the leading exponential time dependence and the associated kinematic factor, and the residual time-dependence was fit to yield $F_{\eta_c}(Q^2)$ for each correlator.
The $Q^2$ dependence of the form factor was fit using the parameterisations in Table~\ref{tab:fit_forms} -- summaries of the results are given in Table~\ref{tab:etace_full_fit_params} in Appendix \ref{sec:ff_tables}.
The extracted $F_{\eta_c}(Q^2)$ are shown as points in Figure \ref{fig:etace} and the bands indicate the results of the fits -- those which have a considerably worse goodness of fit, shown in italics in the table, are excluded from the plot and subsequent analysis.
In all the parameterisations, $F_{\eta_c}(0)$ is a free parameter and its fitted value is unity
within statistical uncertainties as expected
from the renormalisation of the vector current, giving confidence in the results and the determination of $Z_V$.

While the fits appear to reasonably capture the $Q^2$ dependence of the form factor, even the best have $\chi^2/\Ndf \sim 4$.
We note that the data points are statistically precise but with an unquantified systematic uncertainty and this combination may explain the poor goodness of fit. The spread of data points at each $Q^2$ (resulting from different combinations of source and sink momenta) is an indication of lattice artifacts. When values of the form factor determined at larger $Q^2$, corresponding to correlators with $|\vec{p}| \geq 2\tfrac{2\pi}{L}$ at the source, sink and/or insertion, are included in fits, the $\chi^2/\Ndf$ deteriorates further lending support to the argument that lattice artifacts are a significant factor causing this poor goodness of fit.
In agreement with Ref.~\cite{Dudek:2006ej}, for parameterisations with two parameters we find that the VMD-inspired fit form does a poor job of describing the data in contrast to the Gaussian form.

The charge radius, defined by
\begin{align}
  \label{eq:r2_def}
  \left< r^2 \right> = -6 \left. \frac{\mathrm{d}F(Q^2)}{\mathrm{d}Q^2} \right|_{Q^2=0},
\end{align}
can be determined from the $Q^2$ dependence of the form factor at $Q^2=0$ and compared to previous lattice results and other approaches.
To investigate if there is any systematic difference between the charge radius extracted from a spatial or temporal vector current, as well as fits to all the data (Table~\ref{tab:etace_full_fit_params}), we also consider fits involving only data from correlators featuring a temporal insertion and only data from those featuring a spatial insertion -- summaries of the results are given in, respectively, Tables~\ref{tab:etace_temporal_fit_params} and \ref{tab:etace_spatial_fit_params} in Appendix \ref{sec:ff_tables}.
It can be seen that the spread of results from fitting to these subsets of the data is similar in size to the spread from different fit forms.
Taking an envelope over the fits, excluding those shown in italics in the tables, yields,
\begin{align}
  \label{eq:etace_r2}
  \left< r_{\eta_c}^2 \right>^{\frac{1}{2}} = 0.243(7) \, \mathrm{fm} \, ,
\end{align}
where the quoted value and uncertainty have been chosen to encompass the spread over statistical uncertainties from the different fits -- the spread is much larger than the statistical uncertainty
from any given fit and the uncertainty from varying $m_{\eta_c}$ and $\xi_{\eta_c}$ within their uncertainties.

Other lattice studies~\cite{Dudek:2006ej,Chen:2011kpa,Li:2020gau} have found,
\begin{align*}
  \left< r_{\eta_c}^2 \right>^{\frac{1}{2}} =
  0.176(32) \, \mathrm{fm} \,\text{\cite{Dudek:2006ej}}, \:
  0.255(2) \, \mathrm{fm} \, \text{\cite{Chen:2011kpa}}, \:
  0.213(1) \, \mathrm{fm} \, \text{\cite{Li:2020gau}}.
\end{align*}
These results include quenched calculations~\cite{Dudek:2006ej} as well as those with dynamical light quarks ($N_f=2$)~\cite{Chen:2011kpa,Li:2020gau}, compared to this work which has dynamical light and strange quarks ($N_f=2+1$), used different lattice actions and methods, and do not quantify systematic uncertainties, and so a detailed comparison is not appropriate. However, we note that our result is within the range of those values.

Now that we have demonstrated the approach in a calculation of the $\eta_c$ form factor, we will test it further by studying the form factor of an excited pseudoscalar, the $\eta_c'$.

%% file: etace_prime_etace_prime.tex
\subsection{\texorpdfstring{$\eta'_c$}{η'} Form Factor}
\label{sec:etace_prime_ff}

A natural extension of the calculation of the $\eta_c$ form factor is to consider the form factor of the first excited pseudoscalar charmonium meson, the $\eta_c'$.
Correlators involving the $\eta_c'$ are expected to be
statistically noisier than those involving the $\eta_c$ because it is an excited state. In addition,
the reduced symmetry at non-zero momentum means that the $\eta_c'$ can be up to the third excited state in a given lattice irrep\footnote{where the $\chi_{c1}$, $\chi_{c2}$ and $\eta_c$ can all appear with lower energy in the same channel}. Therefore,
a reliable determination of the
form factor relies on the use of appropriate operators in the three-point correlators. These additional challenges mean that it provides a good test of the methodology.
As seen in the example case in Figure \ref{fig:proj_op_plot}, the optimised operators constructed following the methods of Section \ref{sec:2pts} perform well in both extracting a signal for the $\eta_c'$ and minimising contamination from the subleading $\eta_c$ ground state.

\begin{figure}[ht]
  \centering
  \includegraphics[width=\textwidth]{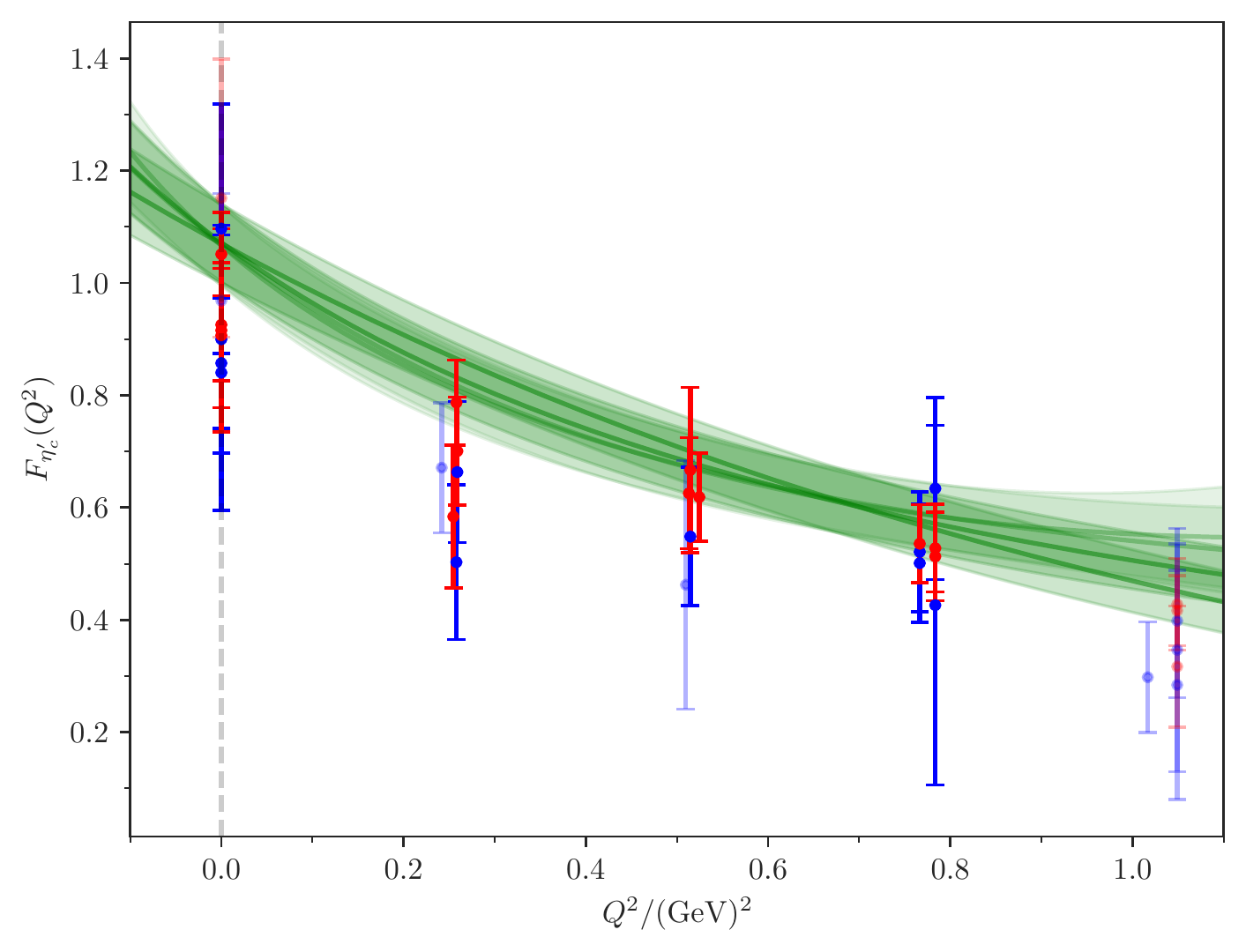}
  \caption{As Figure \ref{fig:etace} but for the $\eta_c'$ form factor.}
  \label{fig:etace_prime}
\end{figure}

The kinematic decomposition for the $\eta_c'$
is identical to that for the $\eta_c$ in eq.~(\ref{eq:etace_decomp}). The analysis proceeds as in Section \ref{sec:etace_ff} and the resulting form factor is shown in Figure \ref{fig:etace_prime} along with results of fits to the $Q^2$ dependence.
Summaries of the fits are given in Table~\ref{tab:etace_prime_etace_prime_full_fit_params} in Appendix \ref{sec:ff_tables}. Additional fits (not shown on the plot) to only data featuring a temporal current or a spatial current are shown in Tables~\ref{tab:etace_prime_etace_prime_temporal_fit_params} and \ref{tab:etace_prime_etace_prime_spatial_fit_params} respectively.
As expected, the data are considerably noisier than for the $\eta_c$ and, probably due to this, the goodness of fits are much better, $\chi^2/\Ndf \sim 1.5$. In all cases $F(0)$ is a free parameter and its fitted value is consistent with unity. The determination of $Z_V^t$ and $Z_V^s$ did not use the $\eta_c'$ and so, in contrast to the $\eta_c$ form factor,
this provides a non-trivial check.

We determine the $\eta_c'$ charge radius in the same way as for the $\eta_c$.
Taking an envelope over fits yields,
\begin{equation*}
  \left< r_{\eta'_c}^2 \right>^{\frac{1}{2}} = 0.62(18) \: \mathrm{fm} .
\end{equation*}
This is larger than the $\eta_c$ charge radius, as one would expect for a radial excitation.
It also has a much larger statistical uncertainty, as expected for an excited state.
In the next section, instead of a radially excited charmonium meson we study an orbital excitation, the scalar $\chi_{c0}$.

%% file: chice_chice.tex
\subsection{\texorpdfstring{$\chi_{c0}$}{χ} Form Factor}
\label{sec:chice_ff}
As for the $\eta_c$, the radiative form factor of the $\chi_{c0}$ is determined by computing correlators with a $\chi_{c0}$ at both the source and sink.
The kinematic decomposition is the same as for the $\eta_c$ in equation (\ref{eq:etace_decomp}) and the analysis proceeds as before. The form factor and results of fits to the $Q^2$ dependence are shown in Figure \ref{fig:chice_overall}.
Summaries of the fits are given in Table~\ref{tab:chice_chice_full_fit_params} in Appendix \ref{sec:ff_tables} -- additional fits (not shown on the plot) to only data featuring a temporal current or a spatial current are shown in Tables~\ref{tab:chice_chice_temporal_fit_params} and \ref{tab:chice_chice_spatial_fit_params} respectively.
The data here are noisier than the ground state $\eta_c$ but less so than for the $\eta_c'$.
In all fits $F(0)$ is a free parameter and is found to be one within statistical uncertainty, giving further confidence in the results and the vector current renormalisation.

Taking an envelope over fits and using the same procedure as for the $\eta_c$, we find the charge radius,
\begin{equation*}
  \left< r^2_{\chi_{c0}} \right>^{\frac{1}{2}} = 0.339(18) \, \mathrm{fm} \, ,
\end{equation*}
Previous lattice studies in the literature~\cite{Chen:2011kpa} (quenched) and \cite{Li:2020gau,Dudek:2006ej} ($N_f = 2$) have found,
\begin{align*}
  \left< r_{\chi_{c0}}^2 \right>^{\frac{1}{2}} =
  0.307(9) \, \mathrm{fm} \, \text{\cite{Dudek:2006ej}}, \:
  0.237(3) \, \mathrm{fm} \, \text{\cite{Chen:2011kpa}}, \:
  0.25(75) \, \mathrm{fm} \, \text{\cite{Li:2020gau}}.
\end{align*}
As previously discussed, results in the literature used different fermion discretisations, different numbers of dynamical flavours and have unquantified systematic uncertainties and so should not be compared in detail.
Nevertheless, we find agreement with \cite{Dudek:2006ej} and the very broad estimate of \cite{Li:2020gau}, though there is some tension with the result in \cite{Chen:2011kpa}.

The hierarchy of charge radii, $ \bigl< r^2_{\eta_{c}} \bigr>^{\frac{1}{2}} < \bigl< r^2_{\chi_{c0}} \bigr>^{\frac{1}{2}} < \bigl< r^2_{\eta_{c}'} \bigr>^{\frac{1}{2}}$, is consistent with expectations from quark-potential models where the $\eta_c$, $\chi_{c0}$, $\eta_c'$ are $nL = 1S$, $1P$, $2S$ states respectively, with $n$ the principal quantum number and $L$ the orbital angular momentum \cite{Dudek:2006ej}.

Now that we have demonstrated the methods in some simple cases, we move to experimentally-accessible transitions from which partial widths can be determined and then more complicated cases involving more than one form factor.

\begin{figure}[ht]
  \centering
  \includegraphics[width=\textwidth]{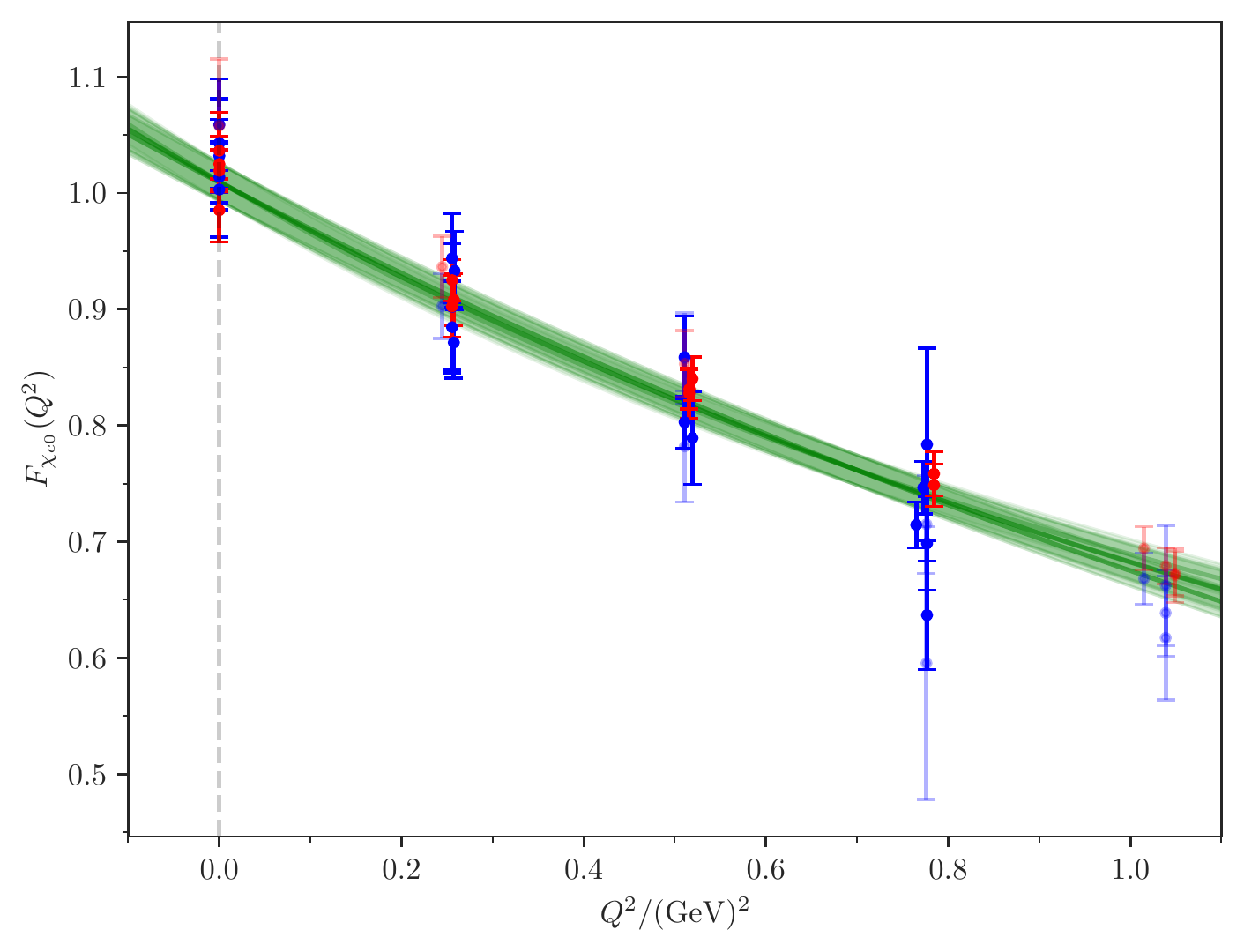}
  \caption{As Figure \ref{fig:etace} but for the $\chi_{c0}$ form factor.}
  \label{fig:chice_overall}
\end{figure}

%% file: psice_etace.tex
\subsection{\texorpdfstring{$J/\psi \to \eta_c$}{J/ψ to η} Transition Form Factor}
\label{sec:psice_etace}

The simplest physical process to study is the $J/\psi \to \eta_c \gamma$ radiative transition for
which the kinematic decomposition is
\begin{equation}
    \begin{split}
        \bra{\eta_c(\pprime)}  j^{\mu}(0) \ket{J / \psi(\lambda, \vec{p} \,)} = \epsilon^{\mu \nu \rho \sigma} p'_{\nu} \, p_{\rho} \, \epsilon_{\sigma} \left(\lambda, \vec{p} \, \right) \frac{2}{m_{J / \psi} + m_{\eta_c}} F_{J / \psi \, \eta_c} \left(Q^2 \right) \, .
        \label{eq:psice_etace_decomp}
    \end{split}
\end{equation}
There is still only a single form factor present but the Lorentz structure is different
from that for the $\eta_c^{(\prime)}$ and involves a polarization vector, $\epsilon_{\sigma} \left(\lambda, \vec{p} \, \right)$, where $\lambda$ is the helicity of the $J/\psi$.

\begin{figure}[ht]
    \centering
    \includegraphics[width=\textwidth]{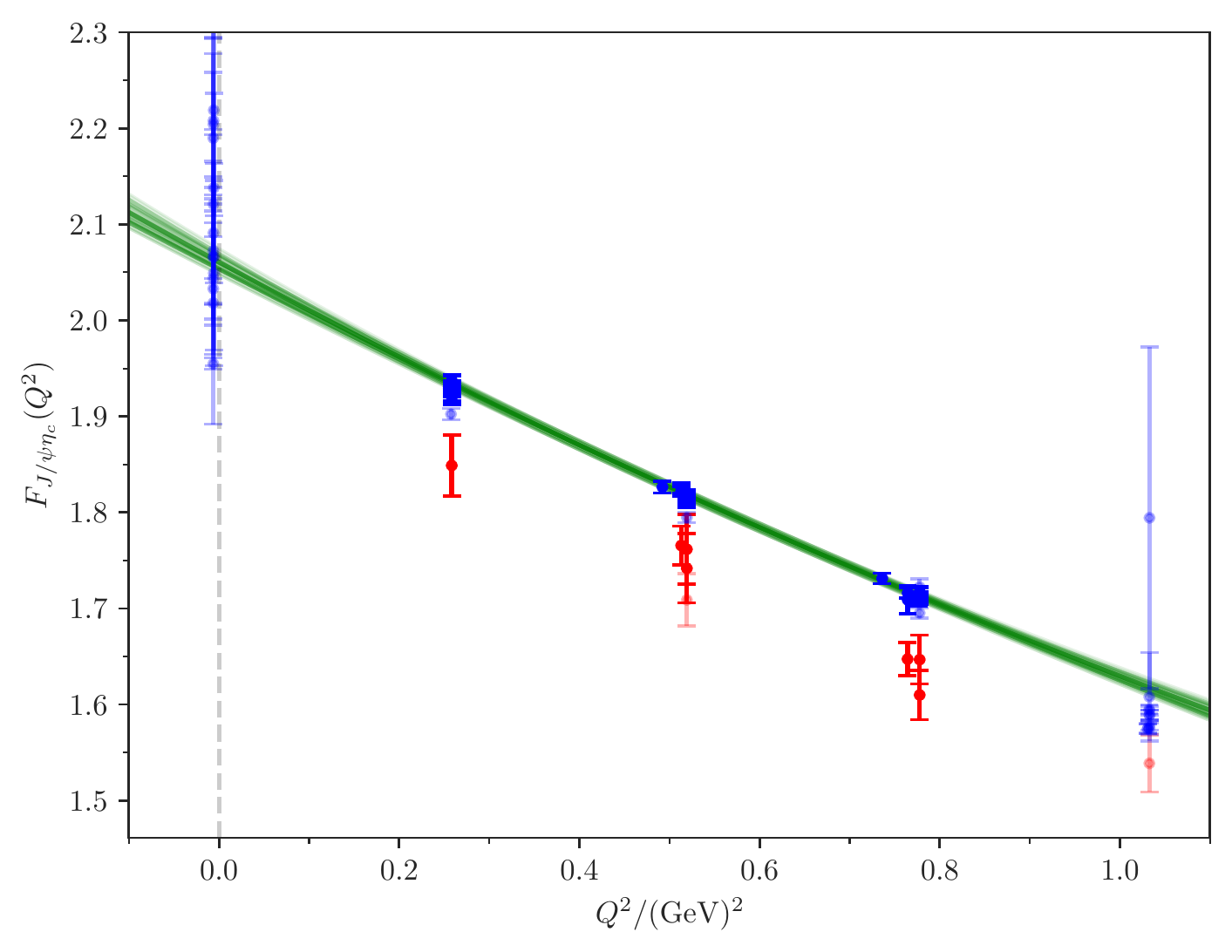}
    \caption{As Figure \ref{fig:etace} but for the $J/\psi \to \eta_c \gamma$ transition form factor.  }
    \label{fig:psice_etace_overall}
\end{figure}

Following the same analysis procedure as above, the resulting form factor is shown in Figure \ref{fig:psice_etace_overall} along with the results of fits to the $Q^2$ dependence.
Summaries of the fits are given in Table~\ref{tab:psice_etace_full_fit_params} in Appendix \ref{sec:ff_tables}. Additional fits (not shown on the plot) to only data featuring a temporal current or a spatial current are shown in Tables~\ref{tab:psice_etace_temporal_fit_params} and \ref{tab:psice_etace_spatial_fit_params} respectively.

The points with small $Q^2$, which correspond to the source and sink having identical momenta,
and one point with $Q^2 \sim 1 \, (\mathrm{GeV})^2$
have much larger error bars than the other points.
For these points the kinematic factor is proportional to $a_t \Delta E = a_t (E_{\eta_c} - E_{J / \psi})$ which is small ($\ll a_t E_{\eta_c}$). These kinematic factors are much smaller than the others and so the signal in the relevant correlators is also much smaller leading to a larger statistical uncertainty on the extracted form factor.
Excluding these points from the fits does not change the results significantly because of their large uncertainties.

The form factor at $Q^2=0$ can be related to the radiative partial width by
\begin{equation}
    \label{eq:psice_etace_width}
    \Gamma(J/\psi \to \eta_c \gamma) = \frac{64\alpha}{27} \frac{|\vec{q}\,|^3}{(m_{J/\psi}+m_{\eta_c})^2} |F_{\Jpsi \, \eta_c}(0)|^2,
\end{equation}
where $\vec{q}$ is the momentum of the photon in the rest frame of the $J/\psi$,
\begin{align}
    |\vec{q}\,|^2=\frac{(m_{J/\psi}^2-m_{\eta_c}^2)^2}{4m_{J/\psi}^2},
\end{align}
which depends strongly on the hyperfine splitting $\Delta m = m_{J/\psi} - m_{\eta_c}$.
There is a discrepancy between the hyperfine splitting measured on this lattice \cite{HadronSpectrum:2012gic} and the experimental value, likely due to it being very sensitive to lattice artifacts.
Therefore, the value of $|\vec{q}\,|$ is dependent on whether lattice or experimental masses are used.
This issue also affects the kinematic factors at small $Q^2$ with large uncertainty discussed above.\footnote{There is a negligible difference between $\Delta m$ and $\Delta E$ when the current has zero spatial momentum.}

One possible approach is to compare $|F_{\Jpsi \, \eta_c}(0)|$ rather than $\Gamma(J/\psi \to \eta_c \gamma)$, avoiding the ambiguity in the choice of $\Delta m$.
In addition, any points with a kinematic factor proportional to the hyperfine splitting can be excluded from the fits, though we find that excluding such points does not change the results significantly because they have large uncertainties.
Following this approach and taking a spread over fit forms we find,
\begin{equation*}
    |F_{\Jpsi \, \eta_c}(Q^2=0)| \sim 1.83 - 2.07.
\end{equation*}
This range of $|F(0)|$
lies at the upper end of the other lattice estimates \cite{Chen:2011kpa, Becirevic:2012dc, Dudek:2006ej, Dudek:2009kk, Gui:2019dtm,Donald:2012ga} but within $1 \sigma$ of most results presented.
Lattice results are consistently larger than the derived PDG value  of $ F_{\Jpsi \, \eta_c}(0) = 1.57(18)$ \cite{Zyla:2020zbs}.
However the most recent CLEO and KEDR results \cite{CLEO:2008pln, Anashin:2014wva} have central values $|F(0)| = 1.62(15), 2.06(11)$, in better agreement with this study.\footnote{All the experimental determinations were calculated by using the provided values of $\Gamma(\Jpsi \to \eta_c \gamma)$ then rearranging eq. (\ref{eq:psice_etace_width}), using the PDG masses.}

All the matrix elements considered thus far have only a single form factor in their decompositions.
In general this is not the case and to test the methods further
we will now consider two channels where multiple form factors occur.

%% file: chice_psice.tex
\subsection{\texorpdfstring{$\chi_{c0} \to J/\psi$}{χ to J/ψ} Transition Form Factors}

The $\chi_{c0} \to J/\psi \, \gamma$ transition presents a new challenge, namely the presence of multiple form factors.
The kinematic decomposition is
\begin{align}
  \bra{ \chi_{c0}(\vec{p}^{\, \prime})} j^\mu(0) & \ket{J/\psi(\vec{p}, \lambda)} = \nonumber                                     \\
  \Omega^{-1}(Q^2)                               & \Bigg( E_1(Q^2) \Big[ \Omega(Q^2)
    \epsilon^\mu(\vec{p}, \lambda) - ( \epsilon(\vec{p}, \lambda) \cdot p' ) \big(
  p^\mu (p \cdot p') - m_{J/\psi}^2 p'^\mu \big) \Big] \nonumber                                                                  \\
  +                                              & C^{\prime}_1(Q^2) \; m_{J/\psi} \; \epsilon(\vec{p}, \lambda) \cdot p^{\prime}
  \Big[ ( p \cdot p') ( p + p' )^\mu - m_{\chi_{c0}}^2 p^\mu - m_{J/\psi}^2 p'^\mu \Big] \Bigg),
  \label{eq:psice_chice_multi_decomp}
\end{align}
where $\Omega(Q^2) = (p \cdot p')^2 - m^2_{J/\psi}m^2_{\chi_{c0}}$.
A modified longitudinal form factor is defined as $C^{\prime}_1(Q^2) = \frac{C_1(Q^2)}{\sqrt{Q^2}}$, where $C_1(Q^2)$ is the usual multipole form factor~\cite{Durand:1962zza, Dudek:2006ej}, to ensure the form factors are real for all $Q^2$.
Note that while the longitudinal form factor $C_1(Q^2 \to 0) \to 0$, the modified form factor $C^{\prime}_1(Q^2)$ does not have this property.
It is important to remember that whilst the presented results will show a non-zero longitudinal $C^{\prime}_1$ form factor at $Q^2=0$, from eq.~\eqref{eq:generic_partial_width} only the electric-dipole form factor, $E_1$, will contribute to the partial width

\begin{equation}
  \Gamma(\chi_{c0} \to \Jpsi \gamma) = \frac{16\alpha}{9} \frac{|\vec{q} \,|}{m^2_{\chi_{c0}}} |E_1(0)|^2.
\end{equation}
As in Section \ref{sec:psice_etace}, there is an ambiguity in whether to use the lattice or experimental meson masses when determining $|\vec{q} \,|$,
but the effect is less pronounced here: for $\Delta m = m_{\chi_{c0}} - m_{J/\psi}$, $\Delta m_{\mathrm{lat.}} = 381.2(1.3) \, \mathrm{MeV}$ and $\Delta m_{\mathrm{exp.}} = 317.8(3) \, \mathrm{MeV}$.

\begin{figure}[htb]
  \centering
  \includegraphics[width=0.89\textwidth]{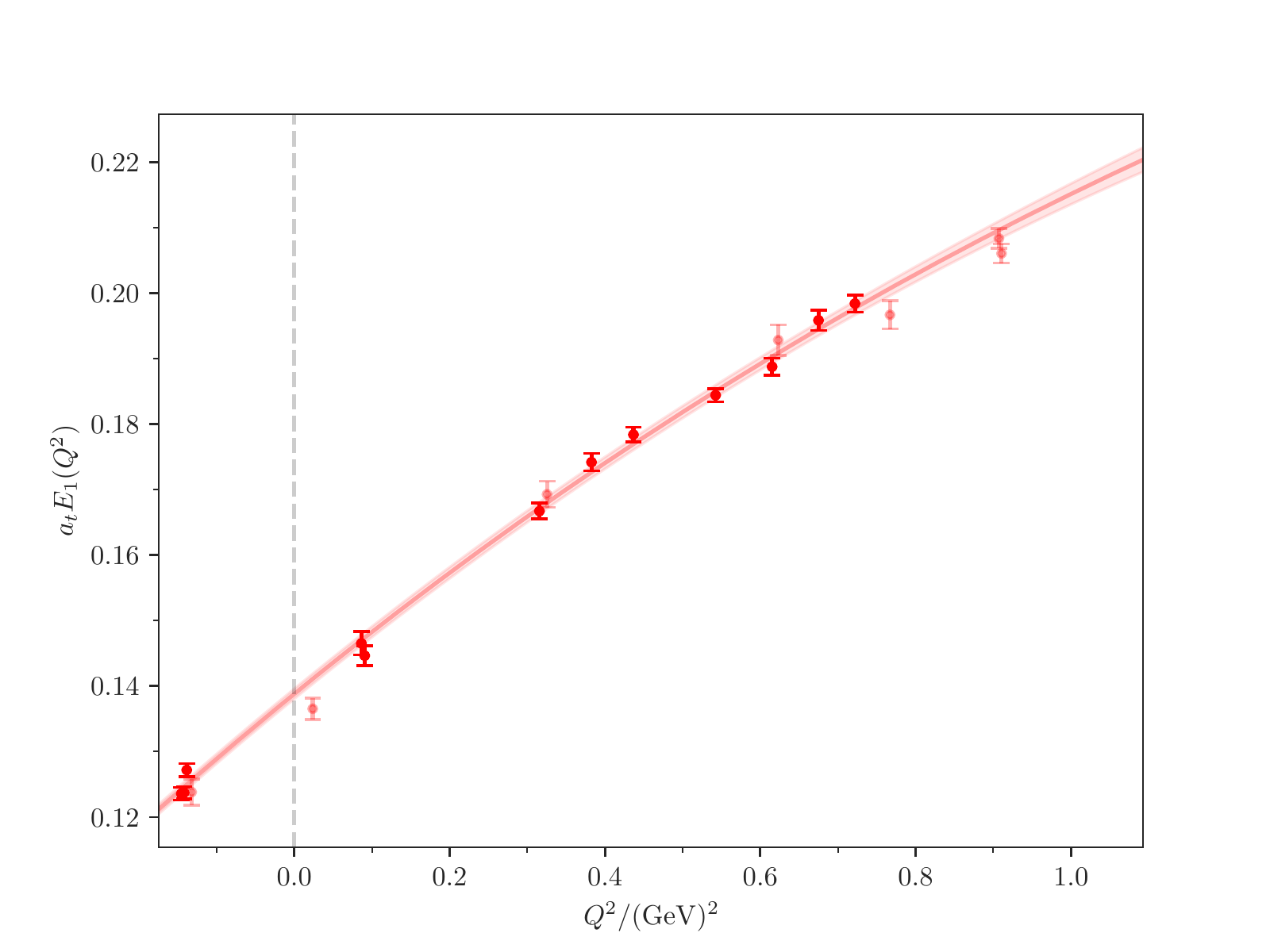}
  \caption{As Figure \ref{fig:etace} but for the $\chi_{c0} \to J/\psi \, \gamma$ electric-dipole form factor, $E_1$.}
  \label{fig:psice_chice_E1}
\end{figure}

\begin{figure}[htb]
  \centering
  \includegraphics[width=0.89\textwidth]{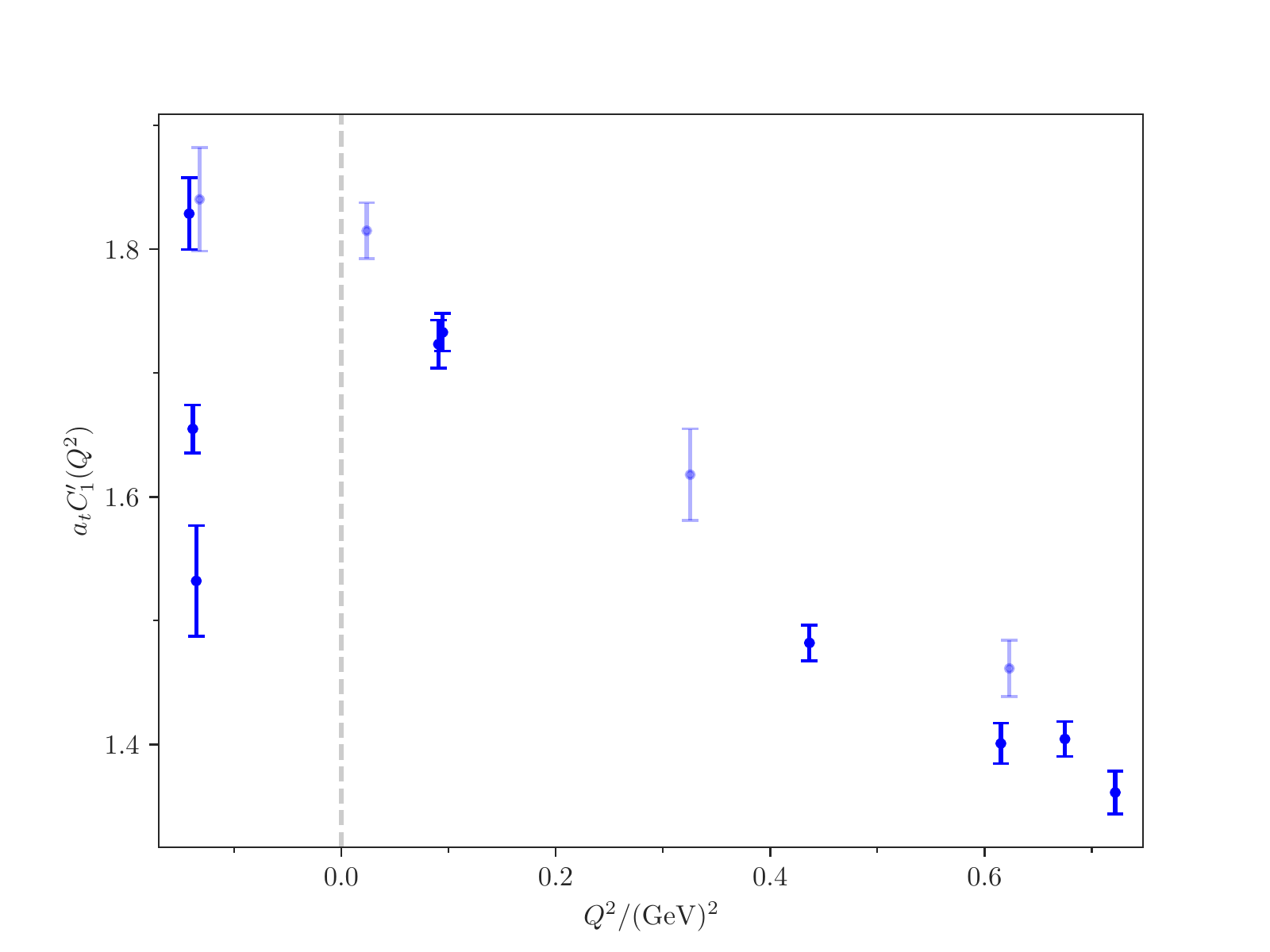}
  \caption{As Figure \ref{fig:psice_chice_E1} but for the modified $\chi_{c0} \to J/\psi \, \gamma$ longitudinal form factor, $C^{\prime}_1$.}
  \label{fig:psice_chice_C1}
\end{figure}

Using the methodology presented in Section \ref{sec:multi_kin}, the resulting $E_1$ and $C^{\prime}_1$ form factors are shown in Figures \ref{fig:psice_chice_E1} and \ref{fig:psice_chice_C1} respectively.
As there is a large tension between points at similar $Q^2$ in Figure \ref{fig:psice_chice_C1}, no fits are performed to the $Q^2$ dependence of $C^{\prime}_1$.
Only the ``Exp2'' parameterisation gives a reasonable description of the $Q^2$ dependence of $E_1$ form factor -- the result is shown in Figure~\ref{fig:psice_chice_E1} and summarised in Table~\ref{tab:psice_chice_E1_fit_params} in Appendix \ref{sec:ff_tables}.
The resulting partial width is,
\begin{equation*}
  \Gamma(\chi_{c0} \to \Jpsi \gamma) = 247.4(2) \, \mathrm{keV}, \quad \Gamma(\chi_{c0} \to \Jpsi \gamma) = 208.6(3) \, \mathrm{keV},
\end{equation*}
using the lattice and experimental masses respectively, where the errors are statistical only.

This is the first determination of $\Gamma(\chi_{c0} \to \Jpsi \gamma)$ with $N_f=2+1$ dynamical quarks.
The result is consistent with \cite{Dudek:2006ej,Dudek:2009kk} (quenched) and \cite{Li:2021gze} ($N_f=2$) from which $\Gamma(\chi_{c0} \to \Jpsi \gamma)$ lies in the range $180-288 \, \mathrm{keV}$. There is some tension with \cite{Chen:2011kpa} ($N_f=2$) who find values in the range $60 - 90 \, \mathrm{keV}$ using a different fermionic action and, like the other results and our result, do not quantify discretization uncertainties.
Furthermore if a comparison is made between $E_1(0)$ values rather than the partial widths, the discrepancy reduces to a factor of approximately $1.5$.

Radiative transitions involving the $\Jpsi$ have been presented here and in the previous section, and the methods have been shown to allow robust determination of multiple form factors. In the next section we apply the same techniques to determine the three form factors of the $\Jpsi$ -- these are unphysical because they violate charge-conjugation parity conservation, but they provide information on the $\Jpsi$ and serve as a further test of the methodology.

%% file: psice_psice.tex
\subsection{\texorpdfstring{$J/\psi$}{J/ψ} Form Factors}

A vector meson has three radiative form factors and the relevant matrix element can be written as
in eq.~\eqref{eq:psice_kindecomp}. The form factors $G_1$, $G_2$ and $G_3$ therein are related to the charge $E_0$, magnetic-dipole $M_1$ and electric-quadrupole $E_2$ form factors by,
\begin{align}
    E_0 & = \left(1 + \tfrac{2 \eta}{3} \right) G_1 - \tfrac{2 \eta}{3} \, G_2 + \tfrac{2 \eta}{3} \left(1 + \eta \right) G_3, \nonumber \\
    M_1 & = G_2,                                     \label{eq:psice_multipole_decomp}                                                   \\
    E_2 & = G_1 - G_2 + \left(1 + \eta \right) G_3,  \nonumber
\end{align}
where $\eta = \tfrac{Q^2}{4m_{\Jpsi}^2}$.

\begin{figure}[tb]
    \centering
    \includegraphics[width=0.89\textwidth]{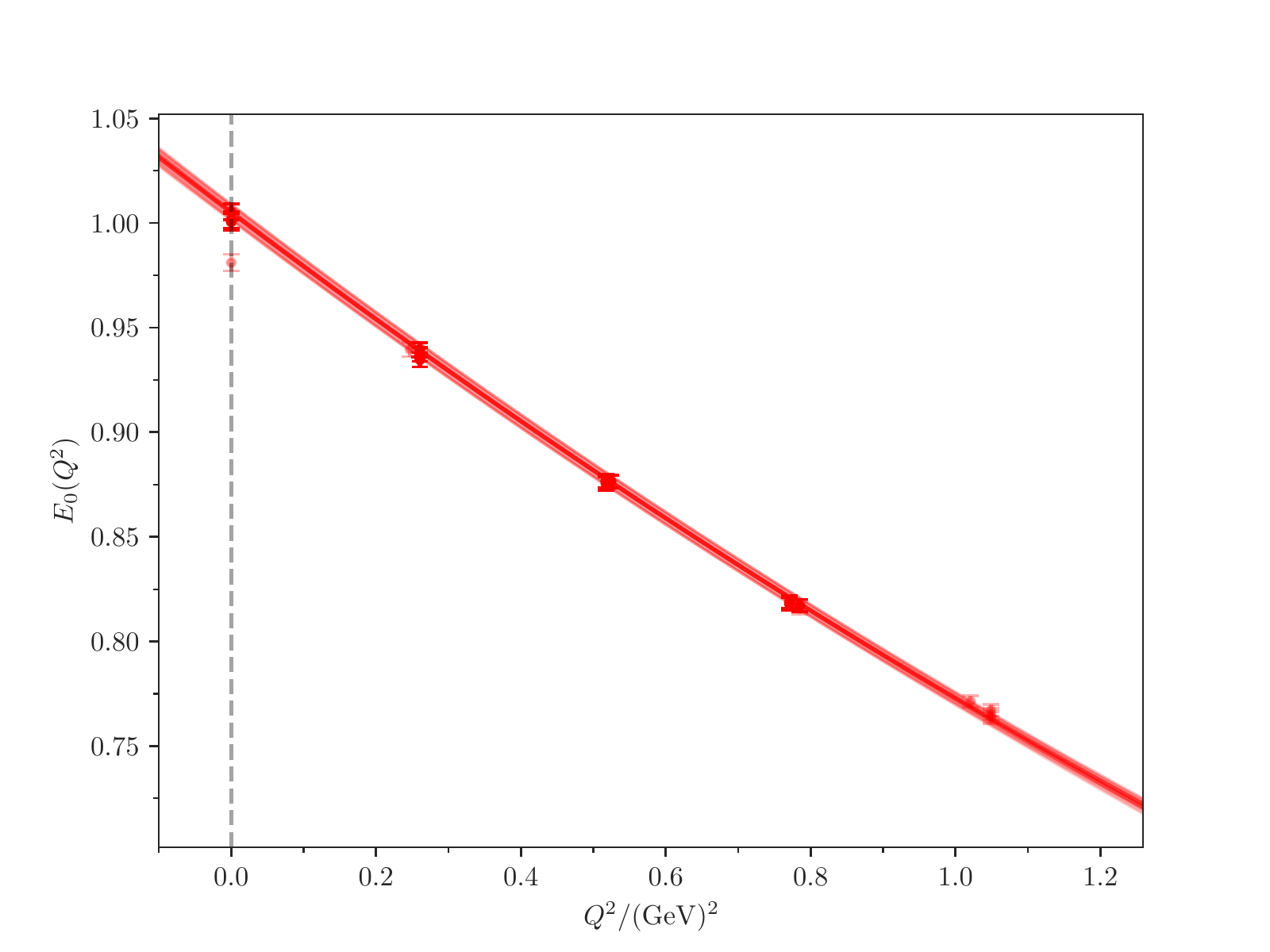}
    \caption{As Figure \ref{fig:etace} but for the $J /\psi$ charge form factor, $E_0$.}
    \label{fig:psice_gc}
\end{figure}

\begin{figure}[tb]
    \centering
    \includegraphics[width=0.89\textwidth]{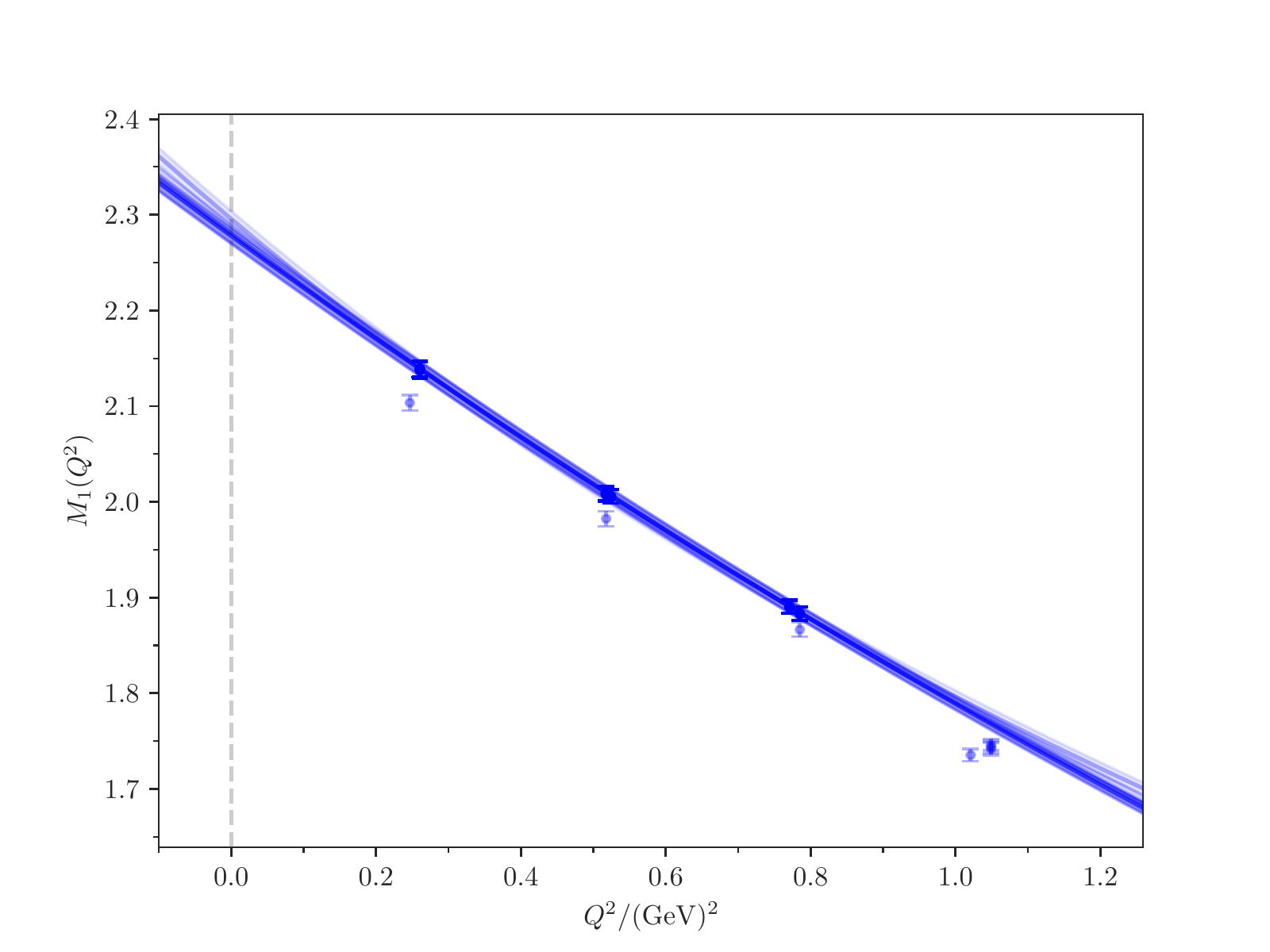}
    \caption{As Figure \ref{fig:etace} but for the $J /\psi$ magnetic-dipole form factor, $M_1$.}
    \label{fig:psice_gm}
\end{figure}

\begin{figure}[tb]
    \centering
    \includegraphics[width=0.89\textwidth]{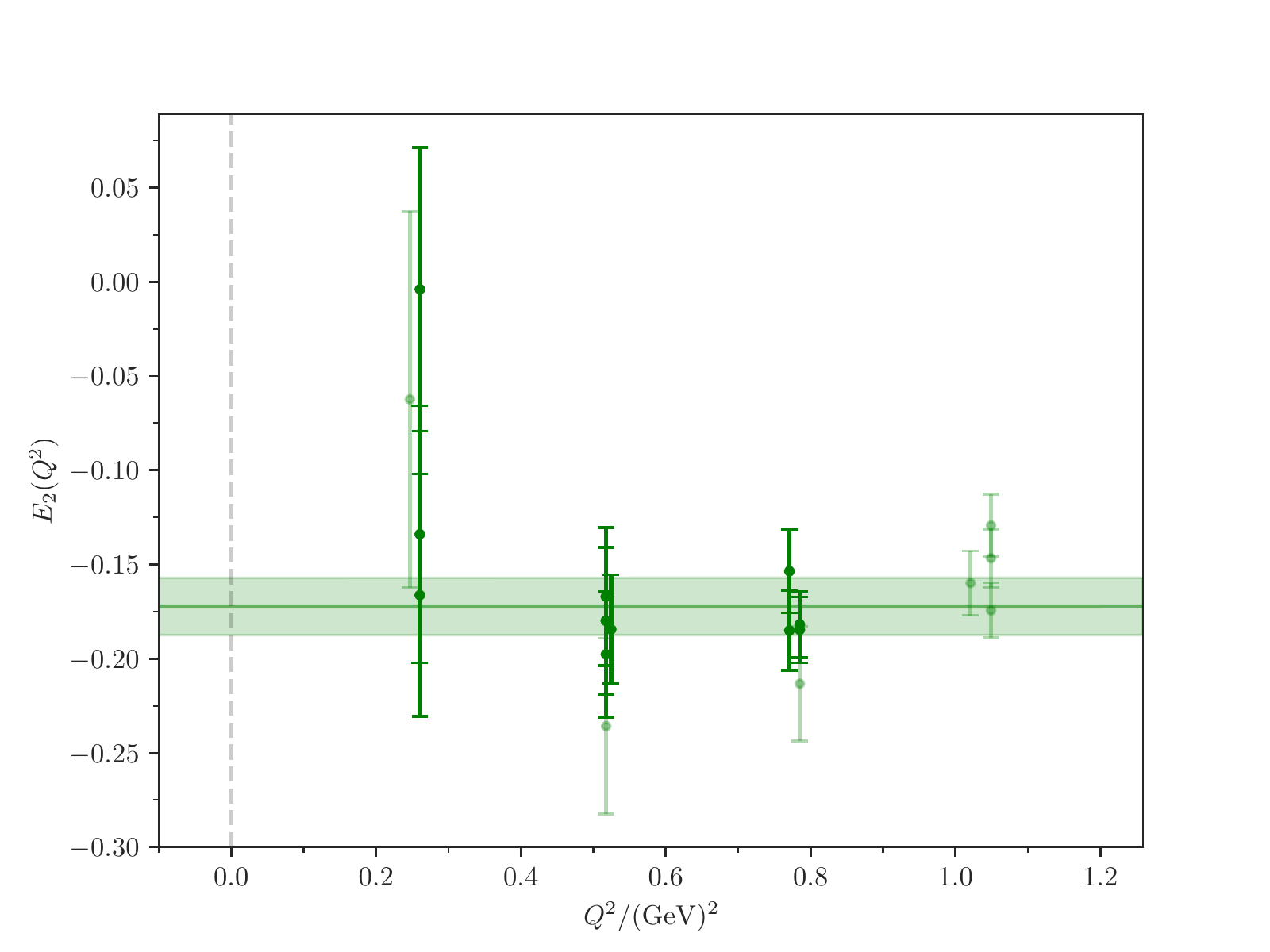}
    \caption{As Figure \ref{fig:etace} but for the $J /\psi$ electric-quadrupole form factor, $E_2$.}
    \label{fig:psice_gq}
\end{figure}

The form factors $E_0(Q^2)$, $M_1(Q^2)$ and $E_2(Q^2)$, determined using the methods outlined above, are shown in Figures \ref{fig:psice_gc}, \ref{fig:psice_gm} and \ref{fig:psice_gq} respectively. Results of fits to the $Q^2$ dependences are also shown in the figures and summaries of the results are given in Tables \ref{tab:psice_E0_fit_params}, \ref{tab:psice_M1_fit_params} and \ref{tab:psice_E2_fit_params} in Appendix \ref{sec:ff_tables}.
Due to the structure of the kinematic factors, direct calculation at $Q^2 \approx 0$ is only possible for $E_0$, while the other form factors rely on an extrapolation in $Q^2$.

For the charge form factor, the $Q^2$ dependence clearly follows a pattern similar to that seen in the $\eta_c$, $\eta_c'$ and $\chi_{c0}$ charge form factors (Figures \ref{fig:etace}, \ref{fig:etace_prime} and \ref{fig:chice_overall}).
We find $E_0(0) = 1.005(4)$, roughly consistent with unity as expected from the vector current renormalisation.
Following the same procedure as for the $\eta_c$ and taking an envelope over the fits gives a charge radius,
\begin{align*}
    \big<r_{\Jpsi}^2 \big>^{\frac{1}{2}} = 0.249(4) \, \mathrm{fm},
\end{align*}
which is comparable to $\big<r_{\eta_c}^2 \big>^{\frac{1}{2}} = 0.243(7) \, \mathrm{fm}$, as would be expected in a quark model as the $\eta_c$ and $J/\Psi$ are both $1S$ states and so have similar spatial wavefunctions.
This result is within the range of previous values in the literature,
\begin{align*}
    \big< r_{\Jpsi}^2 \big>^{\frac{1}{2}} = 0.257(4) \, \mathrm{fm} \: \mathrm{and} \:  0.2570(38) \, \mathrm{fm},
\end{align*}
from \cite{Dudek:2006ej} (quenched) and \cite{Li:2020gau} ($N_f=2$) respectively.

As can be seen in Figure \ref{fig:psice_gm}, the magnetic-dipole form factor, $M_1(Q^2)$, is also robustly determined. Extrapolation to $Q^2 = 0$ yields $M_1(0) = 2.287(18)$, comparable with 2.10(3) found in \cite{Dudek:2006ej}.

In contrast to $E_0(Q^2)$ and $M_1(Q^2)$, the $E_2(Q^2)$ data is much noisier and the $Q^2$ dependence is not clear.
The kinematic factor associated to the quadrupole form factor, $K_{E_2}$, is much smaller than those for the other form factors leading to the magnitude of the contribution to the overall matrix element, $K_{E_2} \, E_2(Q^2)$, being much smaller than the statistical error on the matrix element.
This makes extraction of $E_2(Q^2)$ challenging.
The kinematic factor is larger at larger $Q^2$ improving the quality of the determination -- this can also be seen in \cite{Dudek:2006ej} where the points at
larger $Q^2$ have smaller statistical uncertainties.
Given these large statistical uncertainties, we note that fitting to functions with many parameters is not appropriate.
As such, in Table \ref{tab:psice_E2_fit_params} we only consider fits to a constant, $E_2(Q^2) = E_2(0)$, and two-parameter fit forms.
Taking a spread over them yields \mbox{$E_2(0) = -0.172(15)$}. This is comparable with previous lattice determinations which found $E_2(0) = -0.109(56)$ \cite{Li:2018sfo} and $-0.23(2)$ \cite{Dudek:2006ej}.
Despite the challenges involved, it can be seen that the methods developed in this work can be used to robustly and precisely determine multiple form factors.

%% file: conclusion.tex
\section{Conclusions}
\label{sec:conclusion}

In this study, building on previous work by the Hadron Spectrum Collaboration~\cite{Shultz:2015pfa}, a lattice QCD methodology to compute radiative form factors and transition amplitudes involving charmonia has been developed. The methods have been tested in computations of amplitudes
involving some of the lower-lying charmonia, $\eta_c$, $\Jpsi$, $\chi_{c0}$ and $\eta_c'$.
While we should reiterate that we have used a single lattice spacing and not quantified the various systematic uncertainties,
the results demonstrate that a high level of statistical precision can be achieved, even for excited states, and multiple form factors can be robustly determined.
Whilst the results at $Q^2=0$ have been highlighted, transitions and form factors have been determined at a range of momenta combinations offering insight into the $Q^2$ dependence.
These are the first studies of a number of these amplitudes in a lattice QCD calculation with dynamical light and strange quarks.

Now that the techniques have been demonstrated, in future work they will be applied to study radiative transitions involving higher-lying and exotic charmonia. As has been shown in quenched lattice QCD calculations~\cite{Dudek:2009kk}, studying radiative transitions can provide information on the structure of charmonia and aid in their identification. For example, it will be interesting to study transitions involving hybrid charmonia with $J^{PC}=1^{--}$ and $J^{PC}=1^{-+}$ -- there are suggestions of experimental candidates for the former, but not currently for the latter~\cite{Zyla:2020zbs,Brambilla:2019esw}.
To study charmonia unstable with respect to QCD, the methods can be combined with those developed and demonstrated in Refs.~\cite{Briceno:2015dca,Briceno:2016kkp,Briceno:2017max,Radhakrishnan:2022ubg}.
These calculations are timely given ongoing investigations at BESIII and the LHC, and the planned PANDA experiment.

%% file: normalizing_two_points.tex
\section{Normalisation of Optimised Operators}
\label{sec:overlap_factors}

To extract the relevant matrix elements we compute three-point correlations functions with optimised operators, $\Omega_{\estate{n}}$, at the source and the sink. These variationally-optimal operators are constructed using eq.~(\ref{eq:opt_op_defn}) after solving the generalized eigenvalue problem (GEVP, eq.~(\ref{eq:gevp})). For appropriate $t$ and $t_0$, the eigenvalues, known as principal correlators, will be of the form $\lambda_{\estate{n}}(t) \sim e^{-E_{\estate{n}}(t-t_0)}$. We fit these to the form,
\begin{equation}
    \lambda_\estate{n}(t, t_0) = (1- A_\estate{n})\, e^{- E_\estate{n} (t-t_0) } + A_\estate{n}\, e^{- E'_\estate{n} (t-t_0) } \, ,
    \label{eq:fitform}
\end{equation}
where the second term can account for some excited state contamination and stabilises the fit by allowing us to consider longer time ranges. It is important to note that, to ensure the requirement $\lambda_\estate{n}(t_0, t_0) = 1$ is satisfied, the coefficients of the two exponentials are not independent. For reasonable fits, choices of $t_0$ and fit time ranges, we find $A_\estate{n} \ll 1$ and $E'_\estate{n} > E_{N+1}$, where $E_N$ is the largest energy extracted by the GEVP \cite{Blossier:2009kd} as would be expected if the excited state contamination is small.

The normalisation of these operators must be taken into account when extracting matrix elements.
Inserting two optimised operators $\Omega_\estate{n}(t)$, $\Omega_\estate{m}^\dagger(0)$
into eq.~(\ref{eq:two_point_matrix}) gives,
\begin{align}
    \label{eq:opt_two_point_matrix}
    \bra{0} \Omega_\estate{n}(t) \, \Omega_\estate{m}^\dagger(0) \ket{0} = \sqrt{(2E_\estate{n}) (2E_\estate{m})} \, e^{-t_0 (E_\estate{n} + E_\estate{m})/2 } \, v_i^{(\estate{n})*}  C_{ij}(t)\,  v_j^{(\estate{m})} \, .
\end{align}
Using the GEVP and recalling that the eigenvectors are orthogonal with respect to $C_{ij}(t_0)$, for $\estate{n} = \estate{m}$,
\begin{align}
    \label{eq:prin_corr_side}
    \bra{0} \Omega_\estate{n}(t) \, \Omega_\estate{n}^\dagger(0) \ket{0} & = 2E_\estate{n} \, e^{-t_0 E_\estate{n}} \, \lambda_\estate{n}(t, t_0)                                                                                   \\
                                                                         & = 2E_\estate{n} \left[ (1-A_\estate{n}) \, e^{-E_\estate{n} t} + A_\estate{n} \, e^{-E'_\estate{n} t} \, e^{-(E_\estate{n} - E'_\estate{n})t_0 }\right].
\end{align}
On the other hand, inserting a complete basis of eigenstates in the left hand side of eq.~(\ref{eq:opt_two_point_matrix}) and time-evolving gives,
\begin{align}
    \label{eq:spectral_side}
    \bra{0} \Omega_\estate{n}(t) \, \Omega_\estate{n}^\dagger(0) \ket{0} = \frac{\left|\bra{0} \Omega_\estate{n}(0) \ket{\estate{n}} \right|^2}{2 E_\estate{n}} e^{-E_\estate{n} t}
\end{align}
for large $t$, and so, comparing eq.~\eqref{eq:prin_corr_side} and eq.~\eqref{eq:spectral_side},
\begin{align}
    \label{eq:two_point_norm_factor}
    \left|\bra{0} \Omega_\estate{n}(0) \ket{\estate{n}} \right| = 2 E_\estate{n} \sqrt{1 - A_\estate{n}}.
\end{align}
If $A_{\estate{n}} = 0$, i.e.~there is only one term in eq.~(\ref{eq:fitform}), this reduces to $\left|\bra{0} \Omega_\estate{n}(0) \ket{\estate{n}} \right| = 2 E_{\estate{n}}$.
The plateau value for the diagonal $\Omega_1$ correlator being less than 1 in Fig.~\ref{fig:proj_op_plot} is because $A_{1} > 0$ in that channel and hence $\sqrt{1 - A_1} < 1$.

For a generic three-point correlator with optimised operators, we can expand the double sum in eq.~(\ref{eq:generic_three_point_correlator}),
\begin{equation*}
    C^\mu_{\estate{n} \estate{m}}(\Delta t, t) = \frac{e^{-E_\estate{n}(\Delta t - t)}e^{-E_\estate{m}t}}{4 E_\estate{n} E_\estate{m}} \bra{0} \Omega_\estate{n} (0) \ket{\estate{n}} \bra{\estate{n}} j^\mu (0) \ket{\estate{m}} \bra{\estate{m}} \Omega^\dagger_\estate{m} (0) \ket{0} + \dots
\end{equation*}
where the ellipsis denotes terms suppressed by factors of overlaps onto non-desired states $\bra{\estate{m}'} \Omega^\dagger_\estate{m} (0) \ket{0}$ and $\bra{0} \Omega_\estate{n} (0) \ket{\estate{n}'}$ which are minimised due to the variational method.
As stated in Section \ref{sec:plat_fits} these terms will be small for intermediate time $0 < t < \Delta t$.
Discarding these additional terms then inserting the normalization from eq.~(\ref{eq:two_point_norm_factor}) gives,
\begin{equation*}
    C^\mu_{\estate{n} \estate{m}}(\Delta t, t) = e^{-E_\estate{n}(\Delta t - t)}e^{-E_\estate{m}t} \sqrt{(1-A_\estate{n})(1-A_\estate{m})} \bra{\estate{n}} j^\mu (0) \ket{\estate{m}}.
\end{equation*}
Thus, for each three-point correlator there are two independent normalization factors to be corrected for, one depending on $\estate{n}$ and one on $\estate{m}$, when determining the matrix elements $\bra{\estate{n}} j^\mu (0) \ket{\estate{m}}$.
So for all three-point correlators, the matrix elements are given by,
\begin{equation*}
    \bra{\estate{n}} j^\mu (0) \ket{\estate{m}} =  \frac{e^{E_\estate{n}(\Delta t - t)}e^{E_\estate{m}t}}{ \sqrt{(1-A_\estate{n})(1-A_\estate{m})}} C^\mu_{\estate{n} \estate{m}}(\Delta t, t),
\end{equation*}
where the quantites $E_\estate{n}, E_\estate{m}$ and $A_\estate{n}, A_\estate{m}$ are determined on each configuration via jackknife.

%% file: results_appendix.tex
\section{Form Factor Parameterizations}
\label{sec:ff_tables}

In this appendix we present information about the fits to the $Q^2$ dependence of the form factors using the parameterisations in Table~\ref{tab:fit_forms}. Further details of the fits can be found in the supplemental material. Tables \ref{tab:etace_full_fit_params}, \ref{tab:etace_temporal_fit_params} and \ref{tab:etace_spatial_fit_params} show the fits for the $\eta_c$ form factor including all data, only data extracted from correlators featuring a temporal current insertion and only data extracted from correlators featuring a spatial current insertion respectively. Tables \ref{tab:etace_prime_etace_prime_full_fit_params}, \ref{tab:etace_prime_etace_prime_temporal_fit_params} and \ref{tab:etace_prime_etace_prime_spatial_fit_params} give analogous results for the $\eta_c^\prime$ form factor, Tables \ref{tab:chice_chice_full_fit_params}, \ref{tab:chice_chice_temporal_fit_params} and \ref{tab:chice_chice_spatial_fit_params} results for the $\chi_{c0}$ form factor, and Tables \ref{tab:psice_etace_full_fit_params}, \ref{tab:psice_etace_temporal_fit_params} and \ref{tab:psice_etace_spatial_fit_params} results for the $J/\psi \rightarrow \eta_c \, \gamma$ form factor. Results of fits for the $\chi_{c0} \rightarrow \Jpsi \, \gamma$ $E_1$ form factor are shown in Table \ref{tab:psice_chice_E1_fit_params}, and results for the $\Jpsi$ $E_0$, $M_1$ and $E_2$ form factors are shown in Tables \ref{tab:psice_E0_fit_params}, \ref{tab:psice_M1_fit_params} and \ref{tab:psice_E2_fit_params} respectively.

Any fits that are rejected are shown in italics and are not included in subsequent analyses.  The reason(s) for rejecting the fits are given in the table caption.

\begin{table}[h]
    \centering
    \begin{tabular}{lllrr}
        \hline
        Fit Type & $F_0$      & $\left< r_{\eta_c}^2 \right>^{\frac{1}{2}}$ & $N_{p}$ & $\chi^2 / N_{dof}$ \\
        \hline
        Exp1     & 1.0014(26) & 0.23620(56)                                 & 2       & 8.9                \\
        Exp2     & 1.0014(26) & 0.23620(56)                                 & 3       & 9.4                \\
        Exp3     & 1.0023(26) & 0.2435(13)                                  & 3       & 7.5                \\
        VMD1     & 1.0024(26) & 0.24973(63)                                 & 2       & 8.0                \\
        VMD2     & 1.0023(26) & 0.2442(15)                                  & 3       & 7.5                \\
        ExpVMD   & 1.0023(26) & 0.2442(14)                                  & 3       & 7.5                \\
        \hline
    \end{tabular}
    \caption{Fits to the $Q^2$ dependence of the $\eta_c$ form factor with the fits including data extracted from correlators featuring a temporal or spatial current insertion. For each fit, the form factor at $Q^2=0$, $F_0$, charge radius, $\left< r_{\eta_c}^2 \right>^{\frac{1}{2}}$, number of parameters, $N_p$, and goodness of fit are given, where $\Ndf$ is the number of degrees of freedom.}
    \label{tab:etace_full_fit_params}
\end{table}

\begin{table}[h]
    \centering

    \begin{tabular}{lllrrr}
        \hline
        Fit Type      & $F_0$               & $\left< r_{\eta_c}^2 \right>^{\frac{1}{2}}$ & $N_{p}$    & $\chi^2 / N_{dof}$ \\
        \hline
        Exp1          & 1.0039(26)          & 0.23749(52)                                 & 2          & 5.6                \\
        Exp2          & 1.0039(26)          & 0.23749(53)                                 & 3          & 6.8                \\
        Exp3          & 1.0042(26)          & 0.2386(14)                                  & 3          & 6.6                \\
        \textit{VMD1} & \textit{1.0067(26)} & \textit{0.25274(59) }                       & \textit{2} & \textit{24}        \\
        VMD2          & 1.0043(26)          & 0.2384(15)                                  & 3          & 6.8                \\
        ExpVMD        & 1.0042(26)          & 0.2387(13)                                  & 3          & 6.6                \\
        \hline
    \end{tabular}
    \caption{As Table \ref{tab:etace_full_fit_params} but for fits to the $\eta_c$ form factor involving only data extracted from correlators featuring a temporal current insertion. The fit shown in italics has a considerably worse goodness of fit compared to the others and is excluded from subsequent analysis.}
    \label{tab:etace_temporal_fit_params}
\end{table}

\begin{table}[h]
    \centering
    \begin{tabular}{lllrr}
        \hline
        Fit Type      & $F_0$               & $\left< r_{\eta_c}^2 \right>^{\frac{1}{2}}$ & $N_{p}$    & $\chi^2 / N_{dof}$ \\
        \hline
        Exp1          & 1.0036(27)          & 0.23405(65)                                 & 2          & 4.1                \\
        Exp2          & 1.0036(27)          & 0.23405(66)                                 & 3          & 4.3                \\
        Exp3          & 1.0038(27)          & 0.2371(16)                                  & 3          & 4.1                \\
        \textit{VMD1} & \textit{1.0039(27)} & \textit{0.24751(74)}                        & \textit{2} & \textit{5.9 }      \\
        VMD2          & 1.0038(27)          & 0.2371(18)                                  & 3          & 4.1                \\
        ExpVMD        & 1.0038(27)          & 0.2373(16)                                  & 3          & 4.1                \\
        \hline
    \end{tabular}
    \caption{As Table \ref{tab:etace_full_fit_params} but for fits to the $\eta_c$ form factor involving only data extracted from correlators featuring a spatial current insertion. The fit shown in italics has a considerably worse goodness of fit compared to the others and is excluded from subsequent analysis.}
    \label{tab:etace_spatial_fit_params}
\end{table}

\begin{table}[h]
    \centering
    \begin{tabular}{lllrr}
        \hline
        Fit Type & $F_0$     & $\left< r_{\eta_c^\prime}^2 \right>^{\frac{1}{2}}$ & $N_{p}$ & $\chi^2 / N_{dof}$ \\
        \hline
        Exp1     & 1.070(70) & 0.473(35)                                   & 2       & 1.7                \\
        Exp2     & 1.070(70) & 0.473(35)                                   & 3       & 1.8                \\
        Exp3     & 1.067(70) & 0.596(71)                                   & 3       & 1.6                \\
        VMD1     & 1.073(70) & 0.596(59)                                   & 2       & 1.5                \\
        VMD2     & 1.067(70) & 0.68(12)                                    & 3       & 1.6                \\
        ExpVMD   & 1.072(70) & 0.596(59)                                   & 3       & 1.6                \\
        \hline
    \end{tabular}
    \caption{As Table \ref{tab:etace_full_fit_params} but for the $\eta_c^\prime$.}
    \label{tab:etace_prime_etace_prime_full_fit_params}
\end{table}
\begin{table}
    \centering
    \begin{tabular}{lllrr}
        \hline
        Fit Type & $F_0$     & $\left< r_{\eta_c^\prime}^2 \right>^{\frac{1}{2}}$ & $N_{p}$ & $\chi^2 / N_{dof}$ \\
        \hline
        Exp1     & 1.098(71) & 0.475(35)                                   & 2       & 1.6                \\
        Exp2     & 1.098(71) & 0.475(35)                                   & 3       & 1.8                \\
        Exp3     & 1.096(71) & 0.564(75)                                   & 3       & 1.6                \\
        VMD1     & 1.097(71) & 0.586(58)                                   & 2       & 1.4                \\
        VMD2     & 1.095(71) & 0.62(12)                                    & 3       & 1.6                \\
        ExpVMD   & 1.097(71) & 0.586(58)                                   & 3       & 1.6                \\
        \hline
    \end{tabular}
    \caption{As Table \ref{tab:etace_prime_etace_prime_full_fit_params} but for fits to the $\eta_c^\prime$ form factor involving only data extracted from correlators featuring a temporal current insertion.}
    \label{tab:etace_prime_etace_prime_temporal_fit_params}
\end{table}
\begin{table}
    \centering

    \begin{tabular}{lllrr}
        \hline
        Fit Type       & $F_0$             & $\left< r_{\eta_c^\prime}^2 \right>^{\frac{1}{2}}$ & $N_{p}$    & $\chi^2 / N_{dof}$ \\
        \hline
        Exp1           & 0.83(10)          & 0.387(77)                                   & 2          & 1.1                \\
        Exp2           & 0.83(10)          & 0.387(76)                                   & 3          & 1.2                \\
        \textit{Exp3 } & \textit{0.90(10)} & \textit{0.77(15)}                           & \textit{3} & \textit{0.38 }     \\
        VMD1           & 0.88(11)          & 0.54(14)                                    & 2          & 0.84               \\
        \textit{VMD2 } & \textit{0.90(10)} & \textit{1.06(33)}                           & \textit{3} & \textit{0.37 }     \\
        ExpVMD         & 0.88(11)          & 0.54(14)                                    & 3          & 0.95               \\
        \hline
    \end{tabular}
    \caption{As Table \ref{tab:etace_prime_etace_prime_full_fit_params} but for fits to the $\eta_c^\prime$ form factor involving only data extracted from correlators featuring a spatial current insertion.
        The fits shown in italics are excluded from the analysis because they result in forms that rise dramatically at $Q^2 \gtrsim 1 (\mathrm{GeV})^2$, qualitatively inconsistent with the expected behaviour and the unfitted data near $Q^2 \sim 1 (\mathrm{GeV})^2$. }
    \label{tab:etace_prime_etace_prime_spatial_fit_params}
\end{table}

\begin{table}
    \centering
    \begin{tabular}{lllrr}
        \hline
        Fit Type & $F_0$     & $\left< r_{\chi_{c0}}^2 \right>^{\frac{1}{2}}$ & $N_{p}$ & $\chi^2 / N_{dof}$ \\
        \hline
        Exp1     & 1.009(16) & 0.3140(70)                                  & 2       & 1.6                \\
        Exp2     & 1.009(16) & 0.3140(71)                                  & 3       & 1.6                \\
        Exp3     & 1.010(16) & 0.328(18)                                   & 3       & 1.6                \\
        VMD1     & 1.010(16) & 0.3474(90)                                  & 2       & 1.6                \\
        VMD2     & 1.010(16) & 0.331(23)                                   & 3       & 1.6                \\
        ExpVMD   & 1.010(16) & 0.331(20)                                   & 3       & 1.6                \\
        \hline
    \end{tabular}
    \caption{As Table \ref{tab:etace_full_fit_params} but for the $\chi_{c0}$ form factor.}
    \label{tab:chice_chice_full_fit_params}
\end{table}
\begin{table}
    \centering
    \begin{tabular}{lllrr}
        \hline
        Fit Type & $F_0$     & $\left< r_{\chi_{c0}}^2 \right>^{\frac{1}{2}}$ & $N_{p}$ & $\chi^2 / N_{dof}$ \\
        \hline
        Exp1     & 1.018(17) & 0.3045(81)                                  & 2       & 1.4                \\
        Exp2     & 1.018(17) & 0.3045(82)                                  & 3       & 1.6                \\
        Exp3     & 1.018(17) & 0.316(21)                                   & 3       & 1.6                \\
        VMD1     & 1.017(17) & 0.334(10)                                   & 2       & 1.5                \\
        VMD2     & 1.018(17) & 0.317(26)                                   & 3       & 1.6                \\
        ExpVMD   & 1.018(17) & 0.318(22)                                   & 3       & 1.6                \\
        \hline
    \end{tabular}
    \caption{As Table \ref{tab:chice_chice_full_fit_params} but for fits to the $\chi_{c0}$ form factor involving only data extracted from correlators featuring a temporal current insertion.}
    \label{tab:chice_chice_temporal_fit_params}
\end{table}
\begin{table}
    \centering
    \begin{tabular}{lllrr}
        \hline
        Fit Type & $F_0$     & $\left< r_{\chi_{c0}}^2 \right>^{\frac{1}{2}}$ & $N_{p}$ & $\chi^2 / N_{dof}$ \\
        \hline
        Exp1     & 1.021(20) & 0.337(11)                                   & 2       & 0.76               \\
        Exp2     & 1.021(20) & 0.337(11)                                   & 3       & 0.81               \\
        Exp3     & 1.024(21) & 0.355(33)                                   & 3       & 0.79               \\
        VMD1     & 1.025(20) & 0.381(15)                                   & 2       & 0.77               \\
        VMD2     & 1.024(21) & 0.356(42)                                   & 3       & 0.79               \\
        ExpVMD   & 1.024(21) & 0.357(34)                                   & 3       & 0.79               \\
        \hline
    \end{tabular}
    \caption{As Table \ref{tab:chice_chice_full_fit_params} but for fits to the $\chi_{c0}$ form factor involving only data extracted from correlators featuring a spatial current insertion.}
    \label{tab:chice_chice_spatial_fit_params}
\end{table}

\begin{table}
    \centering

    \begin{tabular}{llrr}
        \hline
        Fit Type & $F_0$      & $N_{p}$ & $\chi^2 / N_{dof}$ \\
        \hline
        Exp1     & 2.0535(59) & 2       & 4.4                \\
        Exp2     & 2.0535(59) & 3       & 4.6                \\
        Exp3     & 2.0604(73) & 3       & 4.5                \\
        VMD1     & 2.0688(60) & 2       & 4.4                \\
        VMD2     & 2.0605(75) & 3       & 4.5                \\
        ExpVMD   & 2.0607(72) & 3       & 4.5                \\
        \hline
    \end{tabular}
    \caption{Fits to the $Q^2$ dependence of the $J/\psi \rightarrow \eta_c \gamma$ form factor with the fits including data extracted from correlators featuring a temporal or spatial current insertion. For each fit, the form factor at $Q^2=0$, $F_0$, number of parameters, $N_p$, and goodness of fit are given, where $\Ndf$ is the number of degrees of freedom.}
    \label{tab:psice_etace_full_fit_params}
\end{table}

\begin{table}
    \centering
    \begin{tabular}{llrr}
        \hline
        Fit Type & $F_0$     & $N_{p}$ & $\chi^2 / N_{dof}$ \\
        \hline
        Exp1     & 1.989(37) & 2       & 0.50               \\
        Exp2     & 1.989(37) & 3       & 0.63               \\
        Exp3     & 1.924(95) & 3       & 0.52               \\
        VMD1     & 2.008(42) & 2       & 0.55               \\
        VMD2     & 1.921(80) & 3       & 0.52               \\
        ExpVMD   & 1.989(38) & 3       & 0.63               \\
        \hline
    \end{tabular}
    \caption{As Table \ref{tab:psice_etace_full_fit_params} but for fits to the $J/\psi \rightarrow \eta_c \gamma$ form factor involving only data extracted from correlators featuring a temporal current insertion.}
    \label{tab:psice_etace_temporal_fit_params}
\end{table}

\begin{table}
    \centering
    \begin{tabular}{llrr}
        \hline
        Fit Type & $F_0$      & $N_{p}$ & $\chi^2 / N_{dof}$ \\
        \hline
        Exp1     & 2.0539(59) & 2       & 3.8                \\
        Exp2     & 2.0539(59) & 3       & 4.0                \\
        Exp3     & 2.0616(73) & 3       & 3.9                \\
        VMD1     & 2.0691(61) & 2       & 3.8                \\
        VMD2     & 2.0619(75) & 3       & 3.9                \\
        ExpVMD   & 2.0620(71) & 3       & 3.9                \\
        \hline
    \end{tabular}
    \caption{As Table \ref{tab:psice_etace_full_fit_params} but for fits to the $J/\psi \rightarrow \eta_c \gamma$ form factor involving only data extracted from correlators featuring a spatial current insertion.}
    \label{tab:psice_etace_spatial_fit_params}
\end{table}

\begin{table}
    \centering
    \begin{tabular}{llrr}
        \hline
        Fit Type         & $a_t E_1(0)$       & $N_{p}$    & $\chi^2 / N_{dof}$   \\
        \hline
        Exp2             & 0.1388(8)          & 3          & 9.0                  \\
        \hline
    \end{tabular}
    \caption{As Table \ref{tab:psice_etace_full_fit_params} but for the $\chi_{c0} \rightarrow \Jpsi \, \gamma$ $E_1$ form factor.}
    \label{tab:psice_chice_E1_fit_params}
\end{table}
\begin{table}
    \centering

    \begin{tabular}{lllrrr}
        \hline
        Fit Type       & $F_0$               & $\left< r_{\Jpsi}^2 \right>^{\frac{1}{2}}$ & $N_{p}$    & $\chi^2 /\Ndf$ \\
        \hline
        Exp1           & 1.0055(36)          & 0.2517(9)                                  & 2          & 4.3            \\
        Exp2           & 1.0049(37)          & 0.2496(23)                                 & 3          & 4.8            \\
        Exp3           & 1.0049(37)          & 0.2496(25)                                 & 3          & 4.9            \\
        \textit{VMD1 } & \textit{1.0097(37)} & \textit{0.2722(10) }                       & \textit{2} & \textit{14}    \\
        VMD2           & 1.0049(37)          & 0.2483(29)                                 & 3          & 4.9            \\
        ExpVMD         & 1.0055(36)          & 0.2517(9)                                  & 3          & 5.0            \\
        \hline
    \end{tabular}
    \caption{As Table \ref{tab:etace_full_fit_params} but for the $\Jpsi$ $E_0$ form factor. The fit shown in italics has a considerably worse goodness of fit compared to the others and is excluded from subsequent analysis.}
    \label{tab:psice_E0_fit_params}
\end{table}

\begin{table}
    \centering
    \begin{tabular}{llrrr}
        \hline
        Fit Type & $F_0$     & $N_{p}$ & $\chi^2 /\Ndf$ \\
        \hline
        Exp1     & 2.278(9)  & 2       & 0.2            \\
        Exp2     & 2.278(9)  & 3       & 0.2            \\
        Exp3     & 2.281(10) & 3       & 0.2            \\
        VMD1     & 2.296(9)  & 2       & 0.9            \\
        VMD2     & 2.281(10) & 3       & 0.2            \\
        ExpVMD   & 2.281(10) & 3       & 0.2            \\
        \hline
    \end{tabular}
    \caption{As Table \ref{tab:etace_full_fit_params} but for the $\Jpsi$ $M_1$ form factor.}
    \label{tab:psice_M1_fit_params}
\end{table}

\begin{table}
    \centering
    \begin{tabular}{llrr}
        \hline
        Fit Type      & $F_0$       & $N_{p}$    & $\chi^2 /\Ndf$ \\
        \hline
        Const         & -0.172(15)  & 1          & 1.5            \\
        Exp1          & -0.172(15)  & 2          & 1.7            \\
        \textit{VMD1} & \textit{--} & \textit{2} & \textit{5.6 }  \\
        \hline
    \end{tabular}
    \caption{As Table \ref{tab:etace_full_fit_params} but for the $\Jpsi$ $E_2$ form factor. As discussed in the text, we only attempt fits to a constant, ``Const'', and the two-parameter ``Exp1'' and ``VMD1'' forms. The fit ``VMD1'' shown in italics has a considerably worse goodness of fit compared to the others and is excluded from subsequent analysis.}
    \label{tab:psice_E2_fit_params}
\end{table}